\pdfoutput=1
\documentclass[%
    aps, 
    physrev,
    reprint,
    floatfix,
]{revtex4-2}

\usepackage{amsmath}
\usepackage{amssymb}
\usepackage{bm}
\usepackage{braket}
\usepackage{comment}
\usepackage[normalem]{ulem}
\usepackage[dvipsnames]{xcolor}

\usepackage{printlen}
\usepackage{float}
\usepackage{graphicx}
\usepackage[table]{xcolor}
\usepackage[caption=false]{subfig}

\usepackage{enumitem}

\usepackage{dcolumn}
\usepackage{multirow}
\usepackage{makecell}
\usepackage{tabularx}
\usepackage{array}
\usepackage{booktabs}
\newcommand{\entry}[2]{\ensuremath{#1~[#2]}}

\usepackage{tikz}
\usepackage{quantikz}
\usepackage[mode=image]{standalone}  %

\makeatletter
\newcount\eqcirc@n
\@for\eqcirc@name:=life_anc,life_even,life_odd,%
    rule_cz_lhs,rule_cz_rhs,rule_prep_lhs,rule_prep_rhs,rule_meas_lhs,rule_meas_rhs,%
    anc_cz1_lhs,anc_cz1_rhs,anc_cz2v_lhs,anc_cz2v_rhs,anc_cz3_lhs,anc_cz3_rhs,%
    prop0a_lhs,prop0a_rhs,prop1a_lhs,prop1a_rhs,prop1c_lhs,prop1c_rhs,prop2a_lhs,prop2a_rhs,%
    life_anc_compact,life_even_compact,life_odd_compact,%
    cprop_mixed_lhs,cprop_mixed_rhs,cprop_both_lhs,cprop_both_rhs\do{%
    \advance\eqcirc@n\@ne
    \expandafter\xdef\csname eqcirc@p@\eqcirc@name\endcsname{\the\eqcirc@n}}
\renewcommand{\eqcirc}[1]{\@ifundefined{eqcirc@p@#1}{\PackageError{main}{No errprop page for '#1'}{}}{%
    \vcenter{\hbox{\includestandalone[scale=0.75, page=\csname eqcirc@p@#1\endcsname]{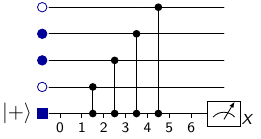}}}}}
\makeatother

\usepackage[colorlinks=true, allcolors=blue]{hyperref}
\usepackage[capitalise]{cleveref}
\usepackage{orcidlink}
\usepackage{tabularx}

\newcommand{\marg}{^\texttt{marg}}
\newcommand{\bnb}{^\texttt{b\&b}}

\newcommand\Yale{
    Departments of Applied Physics and Physics, 
    Yale University, 
    New Haven, CT 06520, USA
    }
    
\newcommand\YQI{
    Yale Quantum Institute, 
    Yale University, 
    New Haven, CT 06511, USA
    }

\newcommand{\Google}{
    Google Quantum AI, 
    Venice, CA 90291, USA
}

\begin{document}

\title{Foliated surface codes for optimal skip-gate leakage correction}

\date{\today}

\author{Runjiang~Bi}
\altaffiliation{These authors contributed equally to this work.}
\affiliation{\Yale}
\affiliation{\YQI}

\author{Nikita~Astrakhantsev}
\altaffiliation{These authors contributed equally to this work.}
\affiliation{\Google}

\author{Thomas~B.~Smith}
\altaffiliation{These authors contributed equally to this work.}
\affiliation{\Yale}
\affiliation{\YQI}

\author{Margaret~Pavlovich}
\affiliation{\Yale}
\affiliation{\YQI}

\author{Ming~Li}
\affiliation{\Google}

\author{Shruti~Puri}
\affiliation{\Yale}
\affiliation{\YQI}

\begin{abstract}
    Skip-gate leakage --- such as atom loss in neutral-atom arrays or trapped ions, or photon loss in superconducting dual-rail qubits --- takes a qubit out of its computational subspace and removes subsequent gates acting on it.
    If leakage is only detectable end-of-the-line at qubit measurement, error correction circuits must be designed to periodically measure all qubits, as in SWAP-based circuits and foliation.
    We show that the best SWAP-based surface code, although optimal against space-like skip-gate leakage errors, is suboptimal against time-like leakage errors.
    By optimal, we mean that the code corrects as many leakage errors as it can detect.
    Remarkably, we find that the foliated surface code is optimal against both space- and time-like error configurations.
    We support our findings with numerical simulations.
    Under two-qubit gate leakage with probability $0.5\%$, foliation suppresses time-like errors in a distance-$4$ rotated stability experiment by $1380(50)$ per two added rounds, compared to $63(1)$ for the best SWAP-based circuit.
    In memory, the two are comparable.
    Under an experimentally motivated neutral-atom noise model, we find that foliation outperforms SWAP-based circuits in stability for both rotated and unrotated code geometries.
    This holds true for a parameter sweep that takes the neutral-atom noise model from being dominated by computational errors to atom loss.
    Interestingly, a SWAP-based circuit for unrotated code memory performs best when computational noise dominates over loss.
    Our work motivates co-designing error correction circuits with logical operations: preferring foliation when time-like errors dominate and SWAP-based circuits when space-like errors dominate.
\end{abstract}

\maketitle

\section{Introduction}\label{sec:introduction}

Leakage of physical qubits outside their computational subspace presents a challenge when implementing quantum error-correcting (QEC) codes.
It manifests as the physical loss of qubits in neutral-atom~\cite{bluvstein2022quantum,bluvstein2025afaulttolerant}, trapped-ion~\cite{coble2025correction,haghshenas2026digital}, and photonic platforms~\cite{bartolucci2023fusion}, and as population transfer from computational levels in superconducting circuits~\cite{barends2014superconducting,rol2019fast,negirneac2021high,mcewen2021removing,miao2023overcoming}.
Left unchecked, leakage can propagate through subsequent gates in a quantum circuit, producing correlated errors that are difficult to correct~\cite{fowler2013coping,ghosh2013understanding,ghosh2015leakage,bultink2020protecting}.

The most effective way to deal with leakage in a QEC code is via non-destructive mid-circuit leakage detection~\cite{wu2022erasure,levine2024demonstrating}.
When interleaved with every circuit operation, mid-circuit detection reveals the exact spacetime location of a leakage error.
In this case, a distance-$d$ code can correct up to $d-1$ leakage errors, while it can only correct $\lfloor (d-1)/2 \rfloor$ Pauli errors~\cite{grassl1997codes}.
Because of this scaling advantage, there is a significant effort towards engineering qubits where the dominant noise is detectable leakage.
Dual-rail qubits in superconducting~\cite{teoh2023dual,kubica2023erasure,mehta2025bias} and photonic~\cite{knill2001scheme} platforms are a leading example, among many others~\cite{ma2023high,scholl2023erasure,kang2023quantum}.

Mid-circuit leakage detection has its drawbacks.
It can be challenging to realize experimentally, adds a new source of noise, and increases the overhead of QEC.
End-of-the-line leakage detection via state-selective measurements is much easier to implement.
State-selective measurements flag leakage, and measure the qubit state in its absence.
These types of measurements have been demonstrated in neutral atoms~\cite{muniz2024high,bluvstein2025afaulttolerant}, trapped ions~\cite{ransford2025helios}, photonics~\cite{lu2008experimental}, and superconducting circuits~\cite{krinner2022realizing,chou2024superconducting}.

When restricted to end-of-the-line leakage detection, a QEC circuit must be designed so that every qubit is periodically measured.
Two common approaches to achieve this are SWAP-based circuits~\cite{suchara2015leakage,brown2019leakage} and foliation~\cite{aliferis2005fault,stace2009thresholds,barrett2010fault,whiteside2014upper,brown2020universal}.
A typical SWAP-based circuit involves the exchange of data and ancilla qubits.
For the surface code, the SWAP can be combined with syndrome extraction, resulting in the walking~\cite{mcewen2023relaxing} and moonwalking~\cite{pavlovich2026erasuresurfacecodecircuit} circuits.
Foliation is implemented by teleporting logical quantum information to new qubits every syndrome extraction cycle~\cite{raussendorf2006fault,raussendorf2007topological,bolt2016foliated,claes2023tailored}.
The teleportation chain removes leakage from the qubits in the code.
The SWAP-based and foliated surface code circuits are expected to have comparable performance under Pauli errors.
However, in this work we show that they behave very differently under leakage.

\begin{table*}
    \centering
    \begin{tabular*}{\textwidth}{
        @{\extracolsep{\fill}}lccccccc@{}
    }       \toprule
        
        & \multicolumn{7}{c}{\textbf{Experiment $(d \times r)$ }}
        \\
        \midrule
        
        & \multicolumn{3}{c}{Memory ($3 \times 3$)}
        & \multicolumn{2}{c}{Stability ($2 \times 3$)}
        & \multicolumn{2}{c}{Stability tCNOT ($2 \times 3$)}
        \\
        
        \cmidrule(lr){2-4}
        \cmidrule(lr){5-6}
        \cmidrule(lr){7-8}
        
        \textbf{Surface code}
        & $d_{\rm L}$
        & $t_{\rm E}\marg$
        & $t_{\rm E}\bnb$
        & $d_{\rm L}$
        & $t_{\rm E}\marg = t_{\rm E}\bnb$
        & $d_{\rm L}$
        & $t_{\rm E}\marg = t_{\rm E}\bnb$ 
        \\
        \midrule
        
        Walking
        & \entry{3}{d}
        & \entry{2}{2d/3}~\cite{pavlovich2026private}
        & \entry{2}{2d/3}~\cite{pavlovich2026erasuresurfacecodecircuit}
        & \entry{3}{r}
        & \entry{2}{r/2}
        & \entry{2}{r} 
        & \entry{2}{r/2}  
        \\
        
        Moonwalking
        & \entry{3}{d}
        & \entry{2}{2d/3}~\cite{pavlovich2026private}
        & \entry{3}{d}~\cite{pavlovich2026erasuresurfacecodecircuit}
        & \entry{3}{r}
        & \entry{2}{r/2}
        & \entry{2}{r} 
        & \entry{2}{r/2} 
        \\
        
        Foliated
        & \entry{3}{d}
        & \entry{2}{2d/3}
        & \entry{3}{d}
        & \entry{3}{r}
        & \entry{3}{r}
        & \entry{3}{r} 
        & \entry{3}{r} 
        \\
        
        \bottomrule
    \end{tabular*}
    
    \caption{
        {\bf Skip-gate leakage error correction for the surface code.}
        The minimum-weight undetectable error is $d_{\rm L}$, and the minimum-weight uncorrectable error using end-of-the-line leakage detection is $t_{\rm E}$.
        We compare SWAP-based circuits (walking~\cite{mcewen2023relaxing} and moonwalking~\cite{pavlovich2026erasuresurfacecodecircuit}) and foliated circuits for the rotated surface code.
        Memory and stability experiments examine the space- and time-like logical errors.
        The space-like dimension $d$ is the code distance, and the time-like dimension $r$ is the number of rounds of stabilizer measurements, as defined in~\cref{sec:experiments}.
        We also study transversal CNOT (tCNOT) gates between two memory or stability surface code patches.
        We calculate $t_{\rm E}$ for two decoders: the branch-and-bound decoder (\texttt{b\&b}) and the marginal decoder (\texttt{marg}).
        The branch-and-bound decoder is slow but accurately accounts for the exclusivity constraint that leakage can only happen once in a qubit's lifetime; the marginal decoder is fast but ignores this constraint~\cite{pavlovich2026erasuresurfacecodecircuit}.
        The quantity in the square bracket is the expected scaling, and the one outside it is the value found via numerical search in a finite-sized circuit whose dimensions are indicated for each experiment.
        The values for $d_{\rm L}$ and $t_{\rm E}$ in a memory tCNOT remain the same as that for a single-patch memory experiment, and are omitted from the table.
    }
    \label{tab:min-weight}
    
\end{table*}

In this paper, we study the \textit{skip-gate leakage} noise model~\cite{liu2026achieving,pavlovich2026erasuresurfacecodecircuit}, in which every gate acting on a leaked qubit is skipped.
This noise model is inspired by neutral-atom~\cite{liu2026achieving}, trapped-ion~\cite{brown2019handling}, and dual-rail qubits~\cite{mehta2025bias}.
It has been shown that under skip-gate leakage, certain SWAP-based circuits can be optimal for surface code memory~\cite{liu2026achieving,pavlovich2026erasuresurfacecodecircuit}.
That is, at least $d$ leakage errors are required to cause an uncorrectable space-like error in a distance-$d$ surface code.
However, we show that these circuits are suboptimal at correcting time-like errors.
For the surface code, we find that the number of leakages required to cause a time-like logical error scales as $\lceil r/2 \rceil$, where $r$ is the number of rounds of stabilizer measurements.
This suboptimal performance results from certain hook errors that can arise because each physical qubit plays the role of data \textit{and} ancilla.
In contrast, we show that the foliated surface code is optimal at correcting both space- and time-like errors.
That is, it can correct up to $d-1$ leakages in space and $r-1$ in time.
Importantly, the role of data and ancilla qubits is never exchanged in the foliated circuit, and the hook errors that arise in the SWAP-based circuits are absent.
As a result, the foliated surface code matches or improves upon the number of correctable leakage errors compared to the SWAP-based alternatives.
This is true for memory, stability~\cite{gidney2022stability}, and transversal CNOT (tCNOT) operations.
These results are summarized in \cref{tab:min-weight}.

Furthermore, we simulate SWAP-based and foliation circuits on a neutral-atom quantum processor under a realistic noise model describing current hardware~\cite{bluvstein2025afaulttolerant}.
Foliation outperforms the SWAP-based circuits in stability for both rotated and unrotated codes.
On the rotated code, SWAP-based circuits also under-perform in memory, largely due to additional noise from moving the $X$ and $Z$ ancilla qubits separately.
On the unrotated code, all ancilla qubits move together, and the moonwalking circuit achieves lower memory error rates than foliation, as the foliated circuit incurs comparatively more idling and trap-transfer errors from teleportation.
No single error correction circuit is best for all logical operations under these experimental constraints.
To get the most out of hardware, error correction and logical circuits should be co-designed to optimize error suppression and lower the overhead of fault tolerance.

This paper is organized as follows.
In~\cref{sec:foliation}, we review the details of foliation.
We identify the critical condition for effectively correcting skip-gate leakage: that there are no weight-1 or weight-2 hook errors.
Then, we demonstrate that the basic foliation circuit for the rotated surface code satisfies this property, as does a more spacetime-compact version.
In~\cref{sec:experiments} we describe the set of experiments used to probe the error suppression of space- and time-like logical errors: memory, stability and tCNOTs.
In~\cref{sec:simulations} we describe a set of numerical simulations.
This includes the exhaustive enumeration of minimum-weight errors and a subthreshold scaling analysis.
We also analyze the performance under a realistic noise model for current neutral-atom processors.
Finally, we conclude in~\cref{sec:conclusions}.

\section{Foliated surface code}\label{sec:foliation}

Here, we describe how the foliated surface code allows for optimal correction of skip-gate leakage errors.
First, in~\cref{subsec:foliation:circuit} we describe the structure of a basic foliation circuit, and study the rotated surface code as an example.
Next, in~\cref{subsec:foliation:min-weight} we argue that a circuit that avoids weight-2 hook errors allows for optimal leakage correction and in~\cref{subsec:foliation:hook-errors} we show that the basic rotated surface code foliation circuit satisfies this requirement.
Finally, in~\cref{subsec:foliation:compact} we describe a lower-depth foliation circuit that preserves the key properties for optimal skip-gate leakage correction.

\begin{figure*}[t]
    \centering
    \includestandalone[width=\textwidth]{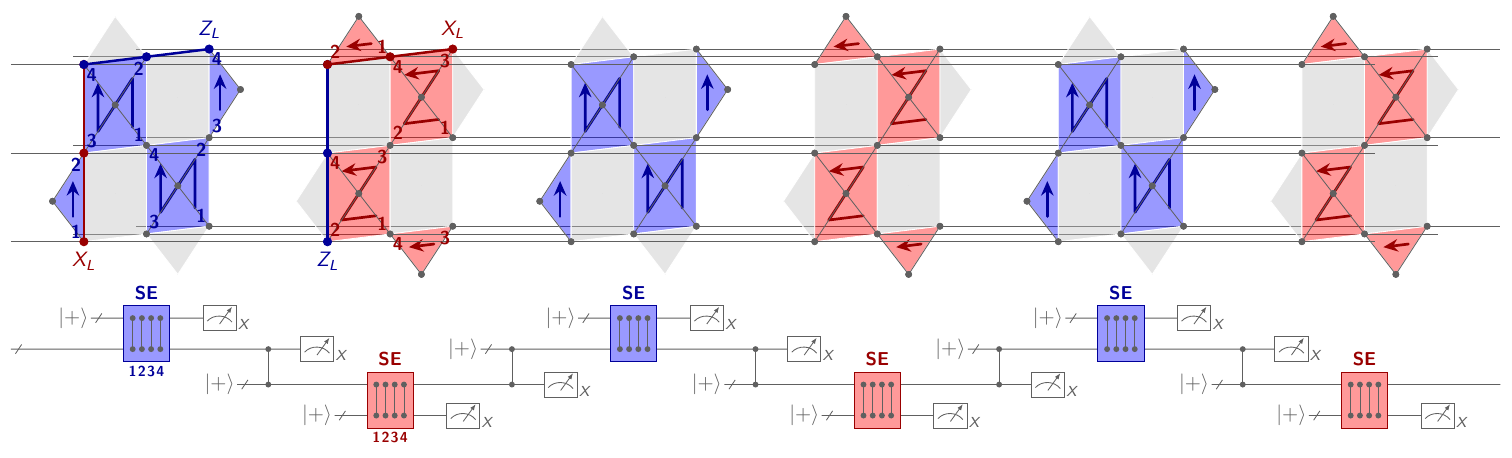}
    \caption{
        {\bf Foliating the rotated surface code.}
        {\it Top:} spacetime circuit for three rounds of stabilizer measurements, where time moves to the right.
        {\it Bottom:} the resulting logical teleportation chain.
        Foliation alternates $Z$ (blue) and $X$ (red) syndrome extraction (SE) with logical teleportation.
        Qubits (gray dots) are prepared in $\ket +$, interact with specific neighboring qubits via CZ gates (gray edges), and then are measured in the $X$ basis.
        Syndrome extraction requires four CZ gate layers, applied in the standard N/Z-ordering to avoid hook errors.
        Logical teleportation involves a single layer of CZ gates, transforming the surface code to its dual; this is reflected in the logical operators $Z_L$ and $X_L$ on one layer being dual to its adjacent layers.
    }
    \label{fig:foliation}
\end{figure*}

\subsection{Basic circuit}\label{subsec:foliation:circuit}

A standard method to foliate a CSS code~\cite{calderbank1996good,steane1996multiple} is to repeatedly measure its stabilizer generators via teleportation~\cite{bolt2016foliated,brown2020universal,claes2023tailored}.
First, the $Z$ stabilizer generators are measured, using an ancilla qubit per stabilizer.
CZ gates are used to map syndrome information from the data qubits to the ancilla qubits.
Then, the logical information is teleported to a new set of qubits via a 1-bit teleportation circuit, again using CZ gates.
As a result of the teleportation step, the code is transformed to its dual, and the $Z$ stabilizers of the dual code (that is, the $X$ stabilizers of the original code) are measured, again using ancilla qubits and CZ gates.
This process is repeated as needed.
\Cref{fig:foliation}~shows the foliation circuit for the rotated surface code for three rounds of $Z$ and $X$ stabilizer measurements.
In this circuit, qubits (gray dots) are prepared in $\ket{+}$, entangled via CZ gates (gray edges), and then measured in the $X$ basis.
To detect leakage errors, every measurement is state-selective.

One round of stabilizer measurements proceeds as follows.
First, the syndrome information of the $Z$ stabilizers (blue) is mapped onto ancilla qubits via four layers of CZ gates.
Measurement of these ancilla qubits reveals the value of the $Z$ stabilizers and thus detects $X$ errors.
After the measurement of $Z$ stabilizers, the data qubits are entangled with a fresh set of qubits via a layer of transversal CZ gates.
Measuring the old data qubits in the $X$ basis teleports the information and applies a transversal $H$, converting the surface code to its dual.
The $Z$ errors in the original code patch are converted to $X$ errors in the new layer.
These errors are detected by measuring the $Z$ stabilizers (red) of the \textit{dual code}, or equivalently the $X$ stabilizers of the original code, by another four layers of CZ gates with ancillas.
After this, the ancilla qubits are measured and the data qubits are teleported onto a new set of qubits using a layer of CZ gates.
This alternate measurement of $Z$ stabilizers in the original code and its dual constitutes one full round of error correction.

Since the CZ gates commute with each other, their order does not affect the measured stabilizer, but it does affect how errors propagate.
In particular, a single error on an ancilla qubit midway through syndrome extraction propagates to correlated $Z$ errors on the data qubits.
To prevent these propagated $Z$ errors aligning with the logical operator $Z_L$, the standard N/Z-gate ordering is used~\cite{tomita2014low}.
This is indicated in~\cref{fig:foliation}.

\begin{figure*}
    \subfloat{\label{subfig:schedule:basic}}%
    \subfloat{\label{subfig:schedule:compact}}%
    \centering
    \includestandalone[width=\textwidth]{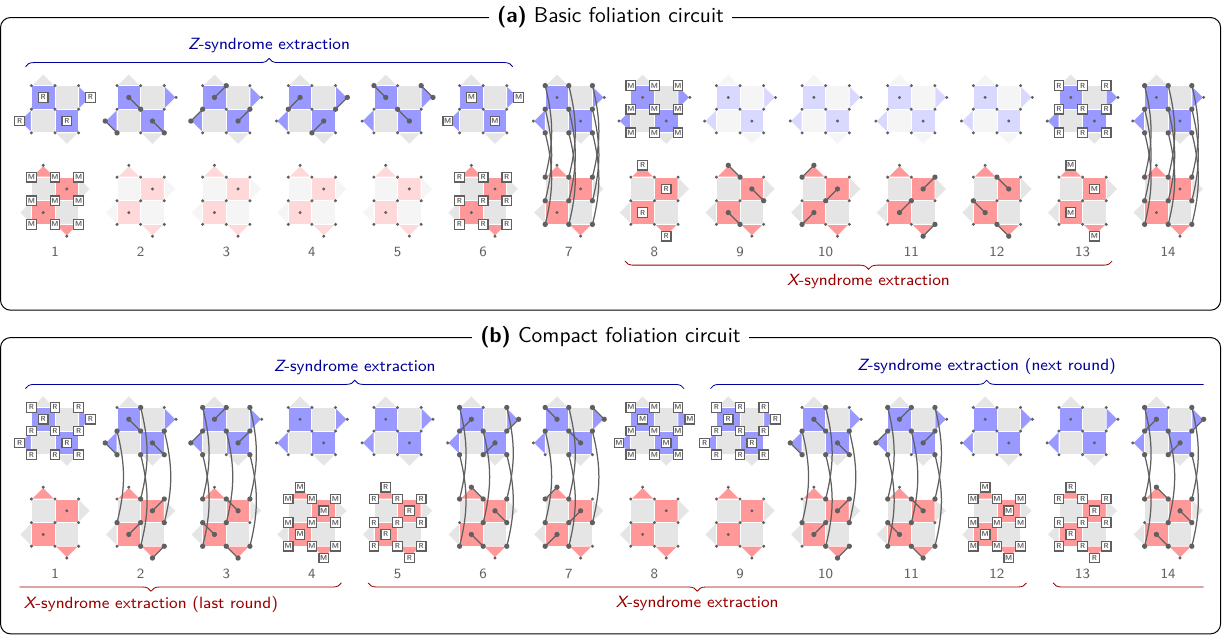}
    \caption{
        {\bf Basic and compact foliation circuits.}
        Boxes {\sf M/R} denote $X$-basis measurement/reset.
        {\bf (a)} {\it Basic foliation circuit:} described in~\cref{subsec:foliation:circuit}.
        Syndrome extraction for $Z$ and $X$ stabilizers occurs sequentially, separated by teleportation layers.
        {\bf (b)} {\it Compact foliation circuit:} described in~\cref{subsec:foliation:compact}.
        Obtained from the basic circuit by moving the teleportation CZ gates into the idling layers.
        The resulting circuit has $X$ and $Z$ syndrome extraction interleaved, though offset in time by half a cycle.
    }
    \label{fig:schedule}
\end{figure*}

The logical performance of the foliated surface code is expected to be comparable to the standard static and SWAP-based syndrome extraction circuits under a uniform depolarizing Pauli noise model.
However, the foliation circuit has two features that make it better suited for leakage correction.
The first is that the total leakage lifetime is short.
By this, we mean that each physical qubit participates in at most four CZ gates between reset and measurement.
Thus, the longest a leakage error can propagate without detection is four gates.
In a SWAP-based surface code circuit it is at most eight gates, since each physical qubit acts as both data and ancilla.
This is valuable when leakage detection is limited to end-of-the-line measurements.
Since the exact spacetime location of leakage is not known, a short leakage lifetime means that there are fewer candidate locations per detected leakage error for the decoder to distinguish.

The second feature is that the foliation circuit contains only CZ gates.
This leads to a lower-weight {\it Pauli envelope} for skip-gate leakage.
The Pauli envelope is a Pauli error model whose effects on a circuit contain the effects of a leakage error, including measurement statistics~\cite{liu2026achieving,pavlovich2026erasuresurfacecodecircuit}.
If the Pauli envelope is decoded correctly, the original leakage errors are guaranteed to be decoded correctly.
A more constrained Pauli envelope is generally better for error correction as it means fewer correlated errors.
As we will show in~\cref{subsec:foliation:min-weight}, the Pauli envelope of a single skip-gate leakage error in the foliation circuit effectively reduces to a depolarization error at the leakage location.
In that section, we use the Pauli envelope formalism to show that the correctability of leakage is determined by the absence of weight-1 and weight-2 hook errors.
We stress that the advantage comes from each qubit playing the same role in each two-qubit gate in the circuit.
That is, they are always a control or always a target.
For the foliation circuit that only uses CZ gates, all qubits only ever play the same role.
Foliation circuits that use a mixture of CX and CZ gates, but still feature `only-target' or `only-control' qubits (such as the generalized foliation variant in Ref.~\cite{claes2023tailored}) will extract a similar benefit, provided that they also satisfy the hook error requirement.
In contrast, a qubit in a SWAP-based circuit changes its role from control to target.
In this case, it is not just weight-1 and weight-2 hook errors, but higher-weight hook errors that will determine correctability.

As shown in~\cref{subfig:schedule:basic}, the basic foliation circuit has a gate depth of 10 (4 CZ gate layers per stabilizer check and 2 CZ gate layers for teleportation).
In~\cref{subsec:foliation:compact} we describe a more compact foliation circuit shown in~\cref{subfig:schedule:compact} that uses only 4 CZ gate layers.
The dense gate structure of the compact circuit is less intuitive, so for clarity in the next two sections we discuss the basic foliation circuit.
The features that allow for optimal leakage correction translate directly to the compact circuit.

\subsection{Minimum-weight uncorrectable errors}\label{subsec:foliation:min-weight}

In this section, we argue that a distance-$d$ foliated CSS code can correct up to $d-1$ skip-gate leakage errors using end-of-the-line leakage detection if its circuit has no {\it weight-1 or weight-2 hook errors}.
We say that a circuit has a weight-$h$ hook error if $h$ errors on the same qubit, between reset and measurement, propagate to a higher-weight error in the support of a logical operator in a way that degrades the overall circuit distance.
Since the foliation circuit for a CSS code involves only CZ gates, we are concerned with the propagation of $X$ errors.
An $X$ error does not commute with the CZ gates and can propagate to a higher-weight error; this is not true for a $Z$ error, which commutes with the CZ gates.

Let $\mathcal L$ denote the set of leaked qubits and $\mathbf t = (t_i)_{i \in \mathcal L}$ the time steps at which each qubit in $\mathcal L$ leaked.
A skip-gate leakage error on qubit $i$ at time $t_i$ can be described by its Pauli envelope $\mathcal P_i(t_i)$~\cite{liu2026achieving,pavlovich2026erasuresurfacecodecircuit}.
For a Clifford circuit composed of CZ, $H$ and $S$ gates, $\mathcal P_i(t_i)$ places an arbitrary Pauli on qubit $i$ at $t_i$, immediately after each subsequent Hadamard on qubit $i$, and before measurement~\cite{liu2026achieving}. 
For CNOT circuits compiled into CZ and $H$, this means a Pauli wherever the qubit changes role from target to control (or vice versa).
Foliated CSS code circuits only use CZ gates, so no qubit undergoes a target-control role change.
The Pauli envelope thus contains only a Pauli at $t_i$ and a Pauli before the $X$-basis measurement.
For the latter, only the $Z$ component affects measurement, and since $Z$ commutes with the CZ gates, it may be absorbed into the Pauli placed at $t_i$.
Therefore, the Pauli envelope of a skip-gate leakage error in foliation is just an arbitrary single-qubit Pauli at the leakage location: $P_i(t_i) \in \{I, X, Y, Z\}$.

A set of leaked qubits $\mathcal L$ produces an error in the set
\begin{equation}\label{eqn:leakage-pauli-envelope}
    \mathcal E_{\mathcal L} (\mathbf t)
    =
    \Big\{
        \prod_{i \in \mathcal L}
        P_i (t_i)
        :
        P_i(t_i) \in \mathcal P_i(t_i)
    \Big\}
    .
\end{equation}
Leakage detection provided by state-selective measurements reveals the set of leaked qubits $\mathcal L$, but not their leakage times $\mathbf t$.
Consequently, the QEC code only needs to separate the spaces corresponding to errors on the same qubit at different times.
Consider two such errors $E \in \mathcal E_{\mathcal L}(\mathbf t)$ and $E' \in \mathcal E_{\mathcal L}(\mathbf t')$.
The code cannot simultaneously correct for both errors if
\begin{equation}\label{eqn:min-weight-condition}
    E E'
    =
    \prod_{i \in \mathcal L}
    P_i (t_i)
    P_i' (t_i')
\end{equation}
forms a nontrivial logical operator.
In the absence of weight-1 and weight-2 hook errors, this condition is only satisfied if $|\mathcal L| \ge d$.

Consider the contribution of a single qubit $i$: its error is $P_i (t_i) P_i' (t_i')$.
Since $Z$ errors commute with the CZ gates, we have $Z_i (t_i) Z_i' (t_i') = 1$, $Z_i (t_i) X_i' (t_i') = Y_i (t_i')$, $X_i (t_i) Z_i' (t_i') = Y_i (t_i)$, and $Y_i (t_i) Y_i' (t_i') = X_i (t_i) X_i' (t_i')$.
Terms that are products of $Y$ and $X$ errors can be written as $ Y_i (t_i) X_i' (t_i') = Z_i (t_i) X_i (t_i) X_i' (t_i')$.
Since the $X$ and $Z$ errors are decoded independently for a foliated CSS code, we only need to consider the combination of a $Y$ and $X$ error as a $Z$ error (say at the beginning of the circuit) and two $X$ errors at times $t_i$ and $t_i'$.
Thus, the only nontrivial errors are then either a single error on the qubit, or a pair of $X$ errors at different time steps: $X_i(t_i) X_i(t_i')$.
If the foliation circuit has no weight-1 or weight-2 hook errors, then in order to get a nontrivial logical operator we need $|\mathcal L| \ge d$.

The standard N/Z-shaped ordering of gates for the rotated surface code is already known to avoid weight-1 hook errors~\cite{tomita2014low,orourke2025compare}.
In the next subsection we show that it also does not have a weight-2 hook error.

\subsection{No weight-2 hook errors in foliated surface code}\label{subsec:foliation:hook-errors}

\begin{figure}
    \centering
    \includestandalone[width=\columnwidth]{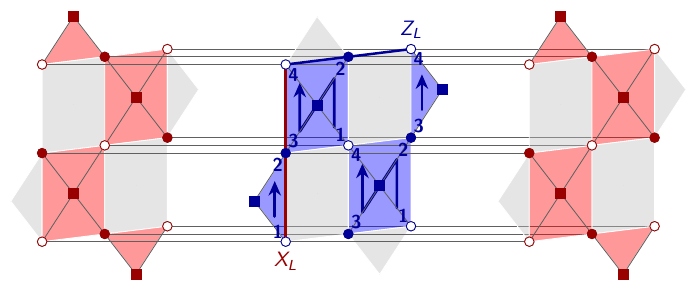}
    \caption{
        {\bf Foliation qubit connectivity.}
        Qubits on the $Z$ stabilizer layer of the foliated rotated surface code.
        Ancilla qubits (blue squares) interact with data qubits (blue circles).
        Data qubits (blue circles) interact with ancilla qubits within the layer, and other data qubits in different layers (red circles).
        Data qubits can be partitioned into two sets: those where the stabilizer check CZs occur sequentially (filled circles), and those where they are separated by idling (empty circles).
    }
    \label{fig:Z-layer}
\end{figure}

To show that the basic surface code foliation circuit has no weight-2 hook errors, we focus on the $Z$-check layer depicted in~\cref{fig:Z-layer}.
In this layer, ancilla qubits (blue squares) interact with four data qubits (blue circles):
\begin{equation}
    \label{eqn:circuit-ancilla}
    \eqcirc{life_anc}
\end{equation}
Each data qubit is teleported in via a layer of CZ gates, interacts with two ancilla qubits, and is teleported out via a final layer of CZ gates.
Depending on its position in the syndrome extraction circuit, a data qubit's two stabilizer check gates can be consecutive (filled circles):
\begin{equation}
    \label{eqn:circuit-data-even}
    \eqcirc{life_even}
\end{equation}
Or, they can be separated by idling (empty circles):
\begin{equation}
    \label{eqn:circuit-data-odd}
    \eqcirc{life_odd}
\end{equation}
We will show that a pair of $X$ errors on any of these qubits during their lifetime does not give rise to a weight-2 hook error.
In our analysis, we use the following identities to determine the propagation of errors:
\begin{subequations}
    \label{eq:rules}
    \begin{align}
        \label{eq:rule-cz} 
        \eqcirc{rule_cz_lhs}
        &= 
        \eqcirc{rule_cz_rhs}
        \\
        \label{eq:rule-prep} 
        \eqcirc{rule_prep_lhs}
        &= 
        \eqcirc{rule_prep_rhs}
        \\
        \label{eq:rule-meas}
        \eqcirc{rule_meas_lhs}
        &= 
        \eqcirc{rule_meas_rhs}
    \end{align}
\end{subequations}
A very similar approach to the one described below can be used to show that there are also no weight-2 hook errors for the qubits in the $X$ stabilizer layer.

\subsubsection{Ancilla qubits}

First we consider hook errors on ancilla qubits.
The circuit for an ancilla qubit is given by~\cref{eqn:circuit-ancilla}.
We classify each configuration of $X$ error pairs according to how many CZ gates there are between them.
In each case, we propagate the first $X$ error forward through the CZs until it annihilates the second $X$ error.
This propagates $Z$ errors onto the data qubits.

If the two $X$ errors are separated by a single CZ, this leaves a single $Z$ error on a data qubit:
\begin{equation}
    \label{eq:anc-cz1}
    \eqcirc{anc_cz1_lhs}
    =
    \eqcirc{anc_cz1_rhs}
\end{equation}
But, if the two $X$ errors are separated by two CZs, this leaves \textit{two} $Z$ errors on separate data qubits.
For example,
\begin{equation}
    \label{eq:anc-cz2v}
    \eqcirc{anc_cz2v_lhs}
    =
    \eqcirc{anc_cz2v_rhs}
\end{equation}
However, due to the gate ordering, the propagated errors are not aligned with any logical operator~\cite{kishony2026surface}.
Thus, the gate ordering successfully avoids weight-2 hook errors due to two $X$ errors separated by two CZs.
If the two $X$ errors are separated by three CZs, this leaves three $Z$ errors, equivalent to a single $Z$ error up to a stabilizer, and the circuit distance is preserved:
\begin{equation}
    \label{eq:anc-cz3}
    \eqcirc{anc_cz3_lhs}
    =
    \eqcirc{anc_cz3_rhs}
\end{equation}
Finally, a pair of $X$ errors separated by all four CZs propagates a weight-4 $Z$ error.
This is just the $Z$ stabilizer itself, which acts trivially on the codespace.

Having described all possible pairs of $X$ errors, we conclude that there are no weight-2 hook errors for ancilla qubits in the rotated surface code.

\subsubsection{Data qubits}

Next we consider hook errors on data qubits.
For simplicity we focus on the circuit in~\cref{eqn:circuit-data-even}.
The process for the circuit in~\cref{eqn:circuit-data-odd} is identical, as the relative gate order is the same.
We again categorize pairs of $X$ errors by how many CZs separate them.

If two $X$ errors are separated by a single CZ, they are equivalent to a $Z$ error on a data qubit in an adjacent check layer.
For example,
\begin{equation}
    \label{eq:data-cz1}
    \eqcirc{prop0a_lhs}
    =
    \eqcirc{prop0a_rhs}
\end{equation}
Clearly, such an error will not be distance reducing.
If two $X$ errors are separated by two CZs, they are equivalent to a pair of $Z$ errors on the partner qubits:
\begin{subequations}
    \begin{align}
        \label{eq:data-cz2-tele}
        \eqcirc{prop1a_lhs}
        &=
        \eqcirc{prop1a_rhs}
        \\
        \label{eq:data-cz2-check}
        \eqcirc{prop1c_lhs}
        &=
        \eqcirc{prop1c_rhs}
    \end{align}
\end{subequations}
The former leads to one space-like and one time-like error and does not reduce the distance of the corresponding logicals.
The latter is a pair of measurement errors on two different time-like logicals and hence is also not distance-reducing.
If the two $X$ errors are separated by three CZs, they are equivalent to a single $Z$ error on a data qubit in a separate layer, which again cannot cause distance reduction.
For example:
\begin{equation}
    \label{eq:data-cz3}
    \eqcirc{prop2a_lhs}
    =
    \eqcirc{prop2a_rhs}
\end{equation}
Finally, a pair of $X$ errors separated by all four CZs can be absorbed entirely by the $X$-basis preparation and measurement and thus has trivial action.

Therefore, we find that there are no weight-2 hook errors for data qubits in the foliated rotated surface code.
This confirms that the foliation circuit satisfies the condition identified in~\cref{subsec:foliation:min-weight} to allow for optimal skip-gate leakage correction.

However, the price for this optimality is circuit depth.
The foliation circuit that we have analyzed thus far requires 10 CZ layers to complete one full round of stabilizer measurements.
In the following section, we present a compactified foliation circuit that interleaves the CZ gates for $Z$ syndrome extraction, $X$ syndrome extraction, and teleportation layers.
We reduce the total depth of a round to 4 CZ layers.

\begin{figure*}
    \subfloat{\label{subfig:experiments:memory}}%
    \subfloat{\label{subfig:experiments:stability}}%
    \subfloat{\label{subfig:experiments:tCNOTs}}%
    \centering
    \includestandalone[width=\textwidth]{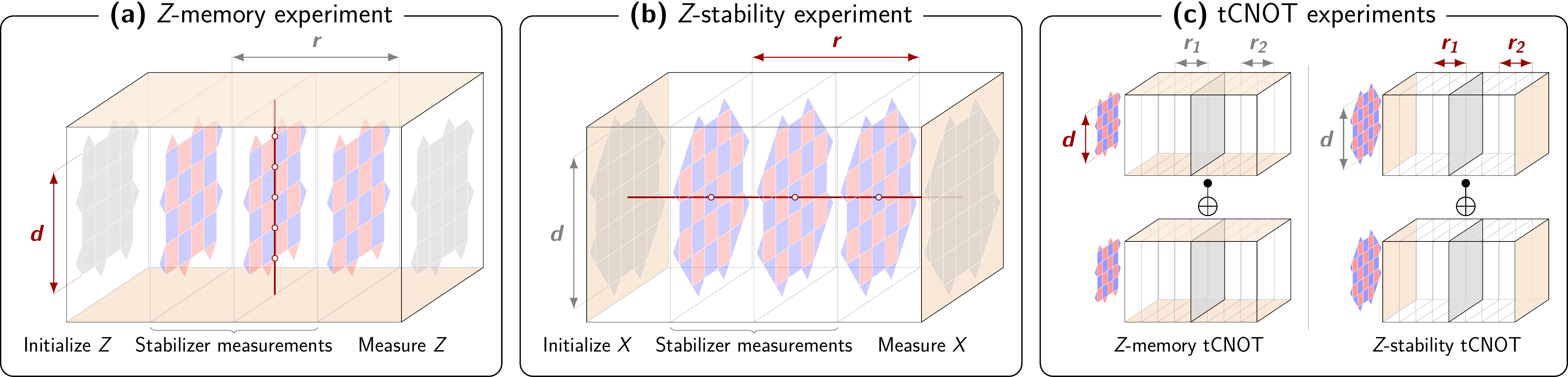}
    \caption{
        {\bf Memory, stability and tCNOT experiments.}
        Spacetime volumes, with time running left to right.
        Shaded faces are open boundaries of the decoding graph, and slices indicate rounds of stabilizer measurements.
        The space-like dimension $d$ is the code distance, and the time-like dimension $r$ is the number of stabilizer measurement rounds.
        {\bf (a)} {\it $Z$-memory experiment:} data qubits are initialized and measured in the $Z$ basis.
        The minimum-weight error (red) is a space-like string of $d$ data qubit $X$ errors.
        {\bf (b)} {\it $Z$-stability experiment:} the patch has $Z$ boundaries all the way around, so data qubits are initialized and measured in the $X$ basis~\cite{gidney2022stability}.
        The minimum-weight error (red) is a time-like string of $r$ ancilla qubit measurement errors.
        {\bf (c)} {\it tCNOT experiments:} a tCNOT (grey layer) between a control and a target patch, with $r_1$ rounds before and $r_2$ rounds after the gate.
        We consider two tCNOT experiments: one with memory patches, and another with stability patches.
    }
    \label{fig:experiments}
\end{figure*}

\subsection{Compact circuit}\label{subsec:foliation:compact}

In this section, we describe a lower-depth foliation circuit for the surface code.
The CZ gates that comprise the circuit all commute with one another, so we are free to perform them in any order, provided that we do not introduce weight-1 or weight-2 hook errors.

Consider physical qubits of the $Z$-check layer in the foliation circuit, shown in~\cref{fig:Z-layer}.
Observe that the ancilla qubit circuit in~\cref{eqn:circuit-ancilla}, and the data qubit circuits in~\cref{eqn:circuit-data-odd,eqn:circuit-data-even} each contain layers of idling.
We compactify the foliation circuit by interleaving the teleportation CZs with the stabilizer check CZs.
For the ancilla qubits, this just means that the idling is removed:
\begin{equation}
    \label{eqn:circuit-ancilla-compact}
    \eqcirc{life_anc}
    \rightarrow
    \eqcirc{life_anc_compact}
\end{equation}
For the data qubits, the teleportation CZs are moved earlier into the idling layers of the circuit:
\begin{subequations}
    \begin{align}
        \label{eqn:circuit-data-even-compact}
        \eqcirc{life_even}
        \rightarrow
        \eqcirc{life_even_compact}
        \\
        \label{eqn:circuit-data-odd-compact}
        \eqcirc{life_odd}
        \rightarrow
        \eqcirc{life_odd_compact}
    \end{align}
\end{subequations}
Only~\cref{eqn:circuit-data-odd-compact} has changed the relative order of the CZ gates; the other two compact circuits only remove idling, so the hook error analysis of~\cref{subsec:foliation:hook-errors} carries over to them unchanged.
For~\cref{eqn:circuit-data-odd-compact}, two $X$ errors separated by one, three or four CZs simplify exactly as before.
For the case of two $X$ errors separated by two CZs, it is still equivalent to a pair of $Z$ errors:
\begin{subequations}
    \begin{align}
        \label{eq:compact-mixed}
        \eqcirc{cprop_mixed_lhs}
        &=
        \eqcirc{cprop_mixed_rhs}
        \\
        \label{eq:compact-both}
        \eqcirc{cprop_both_lhs}
        &=
        \eqcirc{cprop_both_rhs}
    \end{align}
\end{subequations}
As in~\cref{eq:data-cz2-tele,eq:data-cz2-check}, the former is one space-like and one time-like error, and the latter is a pair of measurement errors on two different time-like logicals.
Neither is distance-reducing.

The circuits of the data and ancilla qubits in the $X$-check layer of the foliation circuit can be similarly compactified while also avoiding weight-2 hook errors.
The compactification leads to a dense circuit, where each gate layer interleaves the measurement of the $Z$ and $X$ stabilizer generators, as well as the teleportation gates.
\Cref{fig:schedule} shows the basic and compact foliation schedules.
A single round of stabilizer measurements now requires 4 CZ gate layers, compared with 10 for the basic circuit.

Thus far, we have shown that the foliation circuit for the rotated surface code is optimal for correcting skip-gate leakage errors, as it avoids weight-1 and weight-2 hook errors.
The compact circuit described in this section offsets the increased depth of a single stabilizer measurement cycle while maintaining the resilience to hook errors.
In the following sections, we numerically verify that foliation is capable of optimal correction of both space-like and time-like skip-gate leakage errors.

\section{Memory, stability, and transversal CNOT experiments}\label{sec:experiments}

In this section, we describe the experiments simulated in~\cref{sec:simulations}.
Our goal is to verify that the foliated surface code can optimally correct skip-gate leakage errors in both space and time.
We probe this with memory and stability experiments, respectively.
\Cref{subfig:experiments:memory,subfig:experiments:stability} depict these experiments at a high level. 
Basic and compact foliation circuits for memory and stability are provided in~\cref{app:circuits}, together with a schedule compatible with planar nearest-neighbor architectures.

\Cref{subfig:experiments:memory} shows a memory experiment.
A distance-$d$ code is initialized in the basis of a particular logical operator (in this case $Z$), $r$ rounds of stabilizer measurements are performed, and then the logical operator is measured.
A memory experiment tests how well the logical observables of a code are preserved through time.
It fails if the residual error, after the decoder's correction, forms a space-like string spanning the code patch (shown in red).

\Cref{subfig:experiments:stability} depicts a stability experiment, which is the spacetime dual of memory~\cite{gidney2022stability}.
In a stability experiment, the boundaries of a distance-$d$ code patch are modified so that the product of stabilizers (in this case $Z$) is deterministic, even though the value of each individual stabilizer is not.
Data qubits are initialized in the basis opposite to the boundary type (in this case $X$), the stabilizers of the boundary type (in this case $Z$) are measured $r$ times, and then all data qubits are measured in the basis in which they were initialized.
A stability experiment tests how reliably the product of a set of stabilizers can be determined~\cite{gidney2022stability}.
This is directly relevant to logical two-qubit gates via lattice surgery.
It fails if the residual error, after the decoder's correction, forms a time-like string spanning the circuit (shown in red).

The notable difference between memory and stability experiments is the boundaries of their decoding graphs.
In~\cref{subfig:experiments:memory,subfig:experiments:stability} the shaded boundaries are open, and the unshaded boundaries are closed.
A memory experiment has two open spatial boundaries and closed time boundaries.
A stability experiment has closed spatial boundaries and open time boundaries.
The decoding graphs are effectively spacetime duals of each other.
In~\cref{app:scaling} we use these decoding graphs to analyze the minimum-weight undetectable and uncorrectable leakage errors for each experiment.
 
Stability provides insight into the performance of logical gates via lattice surgery.
However, for reconfigurable architectures like neutral atoms, logical gates can be implemented transversally~\cite{bluvstein2025afaulttolerant}.
It is therefore useful to also probe the performance of a transversal CNOT (tCNOT).
Importantly, a tCNOT gate introduces hyperedges between the decoding graphs of the two code patches, thus affecting the performance of the decoder~\cite{zhou2025low,sahay2025error}.

\Cref{subfig:experiments:tCNOTs} depicts two variants of a tCNOT experiment that we consider in this work.
We refer to these as {\it memory tCNOT} and {\it stability tCNOT}.
The former captures the sensitivity of the tCNOT to space-like logical errors, while the latter captures its sensitivity to time-like logical errors.
In each case, the tCNOT is placed between two surface code blocks, where both blocks are either memory-like or stability-like.
Each code patch has $r_1$ rounds of stabilizer measurements before the tCNOT and $r_2$ rounds after it.
Note that time-like logical errors will not arise when performing global decoding of the full quantum circuit using tCNOT gates because the time boundaries are always closed.
However, they will arise during practical windowed-decoding strategies~\cite{serra2026decoding,turner2026scalable}, where a window's commit boundary acts as an open time boundary.
The stability tCNOT experiment aims to capture the impact of time-like logical errors in this case.
We refer the reader to~\cref{app:tcnot-fault-distance} for more details.

For the foliated surface code, we implement the tCNOT as a tCZ between two transversal $H$ layers on the target patch, where the latter are applied by teleportation.
This adds extra stabilizer measurements on the target patch relative to the control patch.
For this reason, we define $r$ for tCNOT experiments in the following way.
For $Z$-type tCNOTs (i.\,e. a tCNOT between two $Z$-memory or $Z$-stability patches) it is the number of $Z$-stabilizer measurements on the target patch, and for $X$-type tCNOTs it is the number of $X$-stabilizer measurements on the control patch.
The other patch then has $r-1$ $Z$-stabilizer measurements (control) or $r+1$ $X$-stabilizer measurements (target), respectively.
Walking and moonwalking surface code tCNOT experiments have no round-counting issue; in this case both patches have $r=r_1+r_2$ stabilizer measurement rounds.

\subsection{Minimum-weight uncorrectable errors}\label{subsec:experiments:min-weight}

Let $d_{\rm L}$ denote the minimum number of leakage errors that can cause a logical error without triggering any detectors, and let $t_{\rm E} \le d_{\rm L}$ denote the minimum number of leakage errors that can cause an uncorrectable logical error.
The former is a property of the circuit, whereas the latter depends on the decoder.
As explained before, end-of-the-line leakage detection reveals which qubits have leaked but not when, so each qubit has several candidate leakage times.
In this paper, we use the two decoders from Ref.~\cite{pavlovich2026erasuresurfacecodecircuit}.
The branch-and-bound (\texttt{b\&b}) decoder searches for the most likely correction that assigns a single leakage error to each leaked qubit.
The marginal (\texttt{marg}) decoder merges the candidate leakage times into one matching graph and decodes it by minimum-weight perfect matching.
The marginal decoder is much faster than the branch-and-bound decoder.
However, it is less accurate as it does not always produce a correction consistent with a single leakage error on each leaked qubit.

\Cref{tab:min-weight} summarizes the expected scaling of $t_{\rm E}$ with end-of-the-line leakage detection for each combination of experiment, circuit, and decoder.
The predictions for the foliated circuit follow from the analysis in~\cref{sec:foliation}, and are derived in~\cref{app:scaling}.
We expect that, with the branch-and-bound decoder, foliation should exhibit optimal scaling in both space and time: $t_{\rm E}\bnb$ should scale as $d$ in a memory experiment and as $r$ in a stability experiment.
The equivalent predictions for walking and moonwalking memory are obtained from Refs.~\cite{pavlovich2026erasuresurfacecodecircuit,pavlovich2026private}.
Specifically, that $t_{\rm E}\bnb$ for walking is suboptimal and scales as $2d/3$ while the moonwalking circuit is optimal and scales as $d$. 
However, in~\cref{app:walking-stability} we show that both are suboptimal in stability experiments, and both scale as $r/2$.
In~\cref{app:tcnot-fault-distance} we show that the tCNOT experiments follow the same scaling patterns for foliation and moonwalking.
In the following section, we verify these findings numerically.

\section{Numerical simulations}\label{sec:simulations}

In this section, we confirm the predictions of~\cref{sec:experiments} and~\cref{tab:min-weight} with numerical simulations of walking, moonwalking and foliated surface codes.
In~\cref{subsec:simulations:pure-leakage} we analyze the rotated code under gate-only leakage noise.
First, we perform an exhaustive enumeration of the minimum-weight uncorrectable and undetectable errors for small-distance and few-round circuits.
Then, we model higher-distance and higher-round circuits and fit the subthreshold scaling of the logical error rate and failure probability.
In~\cref{subsec:simulations:neutral-atom} we use a realistic noise model for current neutral-atom processors, for both rotated and unrotated codes.

\subsection{Gate-only leakage noise model}\label{subsec:simulations:pure-leakage}

For a baseline theoretical benchmark, in this subsection we study a noise model where leakage occurs at every two-qubit gate with probability $p$.
When leakage occurs, it only affects one of the two qubits, chosen uniformly at random.
We do not include idle errors, or errors due to qubit movement.
Such analysis is deferred to the following subsection.
All numerical simulations described in this subsection are for the rotated surface code.

\subsubsection{Enumerating minimum-weight uncorrectable and undetectable errors}\label{subsubsec:simulations:min-weight}

For each experiment described in~\cref{sec:experiments} we exhaustively search small foliated, walking and moonwalking circuits for the minimum-weight undetectable leakage error $d_{\rm L}$ and the minimum-weight uncorrectable leakage error $t_{\rm E}$.
The $Z$-memory experiments have spacetime dimensions $(d, r) = (3, 3)$ and the $Z$-stability experiments have dimensions $(d, r) = (2, 3)$.
The tCNOT experiments have the same dimensions as their single-patch counterparts; memory tCNOTs have $(d, r) = (3, 3)$ and stability tCNOTs have $(d, r) = (2, 3)$.
For stability tCNOTs we simulate $Z$-stability patches for foliation and $X$-stability for walking and moonwalking.
Recall from~\cref{sec:experiments} that for the foliated $Z$-stability tCNOT, $r$ counts the $Z$-stabilizer measurements on the target block.

The results of our error enumeration are summarized in~\cref{tab:min-weight}.
We omit the results of memory tCNOT from the table as it is identical to the single-patch memory.
As expected, foliation exhibits optimal scaling of $t_{\rm E}\bnb$ for both space- and time-like logical errors.
Furthermore, we find that $t_{\rm E}\marg = t_{\rm E}\bnb$ in stability experiments; we explain this in~\cref{app:mem-stab-scaling:min-weight,app:walking-stability}.

\subsubsection{Subthreshold scaling}\label{subsubsec:simulations:subthreshold}

\begin{figure}
    \subfloat{\label{subfig:subthreshold:memory}}%
    \subfloat{\label{subfig:subthreshold:stability}}%
    \centering
    \includegraphics[width=\columnwidth]{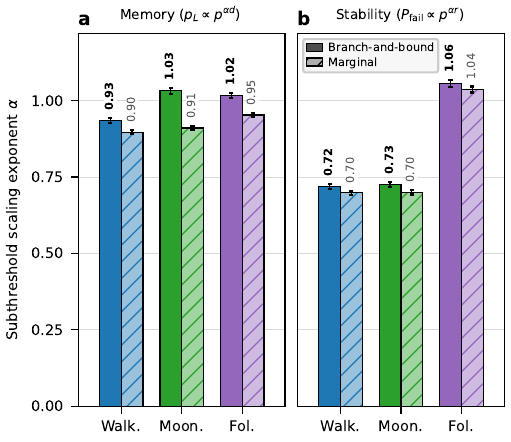}
    \caption{
        {\bf Subthreshold scaling under gate-only leakage.}
        Fitted exponent $\alpha$ for walking, moonwalking and foliated rotated surface code circuits.
        We use both branch-and-bound (solid) and marginal (hatched) decoders.
        {\bf (a)} {\it $X$-memory:} logical error rate per round fit to $p_L = c(p/p_0)^{\alpha d}$.
        {\bf (b)} {\it $Z$-stability:} failure probability per shot fit to $P_{\rm fail} = c(p/p_0)^{\alpha r}$.
        Error bars are likelihood-ratio intervals within a factor of 9 of the maximum likelihood ($96\%$ confidence intervals).
    }
    \label{fig:subthreshold}
\end{figure}

For further verification, we simulate larger circuits and extract their subthreshold scaling.
We simulate $X$-memory experiments with distances $d \in \{5, 7, 9\}$ and rounds $r = 4d$, and fit the logical error rate per round to $p_L = c(p/p_0)^{\alpha d}$.
We simulate $Z$-stability experiments with code distance $d = 4$ and rounds $r \in \{3, 5, 7\}$, and fit the failure probability per shot to $P_{\rm fail} = c(p/p_0)^{\alpha r}$.
In both cases, $c$ and $p_0$ are independent of $d$ and $r$.

\Cref{fig:subthreshold} shows the fitted exponents.
Foliation matches the predicted scaling in~\cref{tab:min-weight}.
Specifically, with the branch-and-bound decoder, we obtain $\alpha \approx 1$ for both memory and stability experiments, which is indicative of optimal suppression of both space- and time-like logical errors.
We also observe optimal space-like error suppression from moonwalking memory, which is consistent with Ref.~\cite{pavlovich2026erasuresurfacecodecircuit}.
Logical error rates, failure probabilities and subthreshold fits are provided in~\cref{app:circuits}.

\subsection{Neutral-atom noise models}\label{subsec:simulations:neutral-atom}

\begin{figure}
    \subfloat{\label{subfig:loss_only_comparison:atom_loss}}%
    \subfloat{\label{subfig:loss_only_comparison:swap_partners}}%
    \subfloat{\label{subfig:loss_only_comparison:cz}}%
    \subfloat{\label{subfig:loss_only_comparison:handoff}}%
    \centering
    \includegraphics[width=\columnwidth]{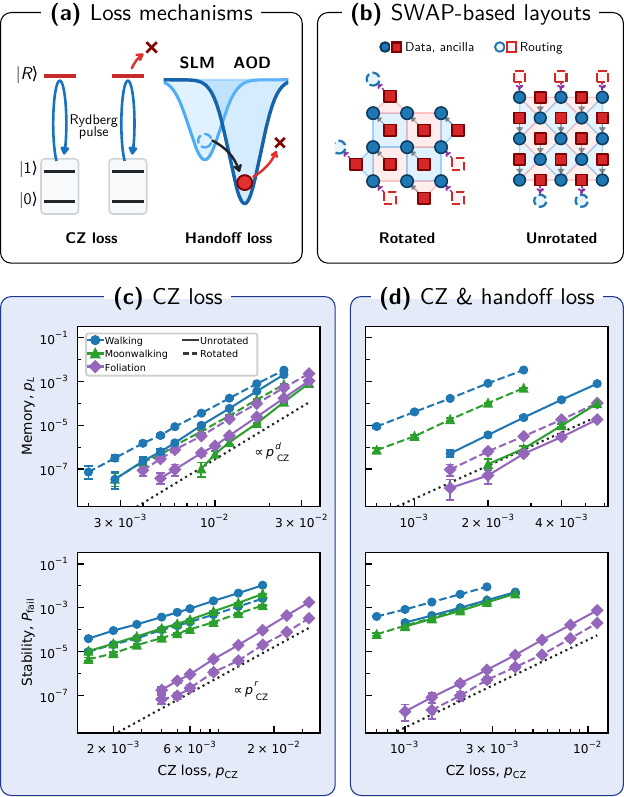}
    \caption{
        {\bf Memory and stability under atom loss.}
        {\bf (a)} {\it Loss mechanisms:} atom loss occurs during a CZ gate and during a handoff between static SLM and mobile AOD traps.
        {\bf (b)} {\it SWAP-based layouts:} distance-$3$ rotated and unrotated surface codes.
        Arrows connect each qubit to its partner; dashed squares and circles are routing qubits.
        {\bf (c)} {\it CZ loss:} $Z$-memory logical error rate per round $p_L$, and $Z$-stability failure probability $P_{\rm fail}$.
        {\bf (d)} {\it CZ \& handoff loss:} the same, but with handoff loss $p_{\rm handoff} = 2p_{\rm CZ}$.
        Dotted lines indicate optimal scaling.
        For both we use the branch-and-bound decoder.
        Error bars are $96\%$ confidence intervals.
    }
    \label{fig:loss_only_comparison}
\end{figure}

In this subsection, we simulate walking, moonwalking, and foliation performance for a neutral-atom quantum processor.
We decompose the circuits into native operations, including atom movement.
Since the required movement depends on the geometry, we consider both the rotated and unrotated surface codes.

In a neutral-atom array, atoms are held either in static spatial light modulator (SLM) traps or in mobile acousto-optic deflector (AOD) traps.
The primary source of leakage is atom loss.
As shown in~\cref{subfig:loss_only_comparison:atom_loss}, atom loss typically occurs either during the Rydberg pulse that drives the native CZ gate, or during a handoff between SLM and AOD traps.
In our simulations, each atom in a CZ gate is lost independently with probability $p_{\mathrm{CZ}}/2$ before the gate, and during a handoff with probability $p_{\mathrm{handoff}}$.
We assume this loss is detected perfectly at end-of-the-line state-selective measurements.

For walking and moonwalking, each round measures all $X$ and $Z$ stabilizers in 4 CZ steps.
The data-ancilla role-swap is merged into the last step (walking) or the first step (moonwalking), and the CZ order is reversed every other round.
This causes the surface code patch to move back and forth between two positions.
Every boundary data qubit needs a role-swap partner.
With a single AOD, we require extra routing qubits to act as these swap-partners, as shown in~\cref{subfig:loss_only_comparison:swap_partners}.
On the rotated code, the AOD moves the $X$- and $Z$-ancilla qubits one after the other.
Each step therefore takes 2 CZ layers, giving 8 CZ layers and about 16--20 handoffs per data qubit per round.
On the unrotated code, $X$- and $Z$-ancilla qubits move together, giving 4 CZ layers and about 7--9 handoffs per data qubit per round.
For foliation, we simulate the basic circuit described in~\cref{subsec:foliation:circuit} (on both rotated and unrotated geometries), with 10 CZ layers per round and about 10 handoffs per data qubit per round.

In~\cref{app:partition_function} we describe a path-counting model that helps interpret the results of our simulations.
Specifically, it explains the reduced stability scaling of walking and moonwalking under pure CZ loss.

\subsubsection{CZ loss and handoff loss}\label{subsubsec:simulations:circuit-level}

\begin{figure*}
    \centering
    \includegraphics[width=\textwidth]{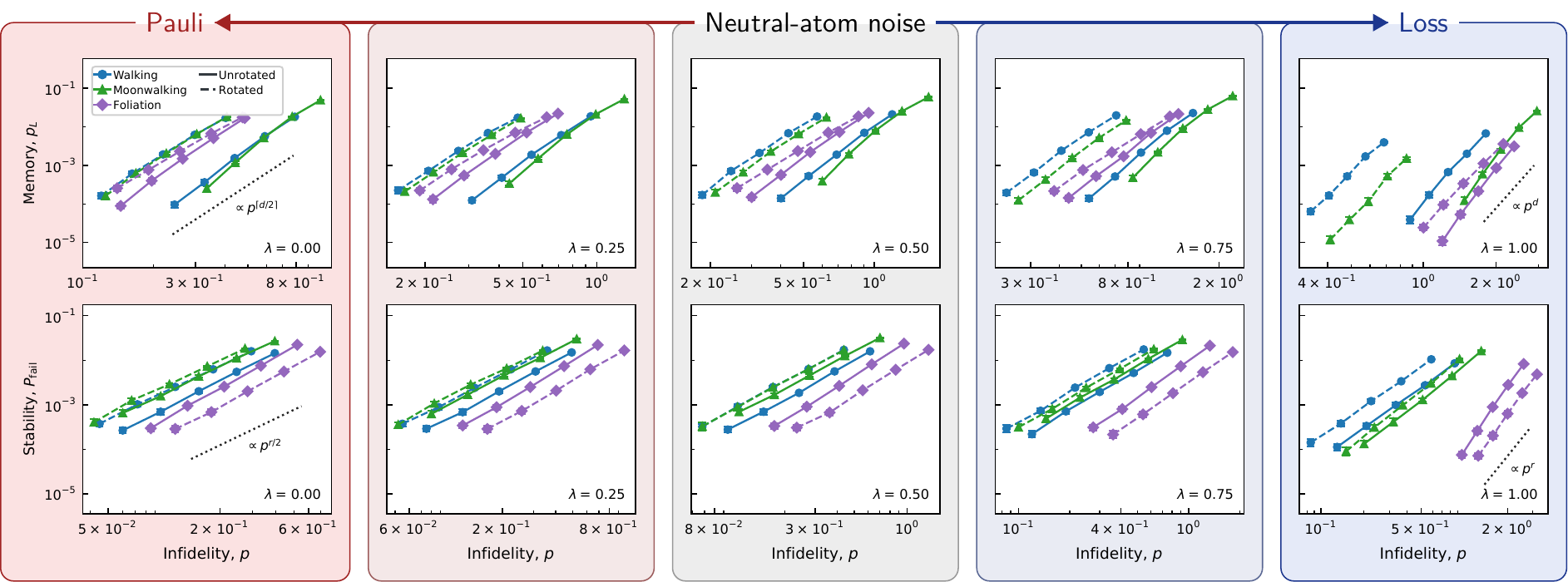}
    \caption{
        {\bf Memory and stability under realistic neutral-atom noise.}
        {\it Top row:} $Z$-memory logical error rate per round $p_L$.
        {\it Bottom row:} $Z$-stability failure probability $P_{\rm fail}$.
        The central panel corresponds to the realistic neutral-atom noise model described in~\cref{app:harvard_noise_model}.
        A parameter $\lambda$ sweeps this noise model from being dominated by Pauli errors ($\lambda=0$, left) to loss ($\lambda=1$, right).
        Dotted lines indicate the subthreshold scaling $\alpha = 0.5$ (left) and $\alpha = 1$ (right).
        We use the branch-and-bound decoder.
        Error bars are $96\%$ confidence intervals.
    }
    \label{fig:realistic_noise_interpolation}
\end{figure*}

\begin{table}
    \centering
    \begin{tabular*}{\columnwidth}{@{\extracolsep{\fill}}lcccc}
        \toprule
        & \multicolumn{2}{c}{Memory} & \multicolumn{2}{c}{Stability} \\
        \cmidrule(lr){2-3}
        \cmidrule(lr){4-5}
        \textbf{Surface code} & Rotated & Unrotated & Rotated & Unrotated \\
        \midrule
        Walking     & $36(1)$  & $65(2)$   & $33(1)$    & $21.0(4)$ \\
        Moonwalking & $83(3)$  & $490(40)$ & $63(1)$    & $53(1)$ \\
        Foliated    & $110(3)$ & $310(10)$ & $1380(50)$ & $900(40)$ \\
        \bottomrule
    \end{tabular*}
    \caption{
        {\bf Error suppression, CZ loss.}
        Suppression factor $\Lambda$ for walking, moonwalking, and foliated circuits on the rotated and unrotated surface code.
        Evaluated for a CZ loss probability of $p_{\mathrm{CZ}} = 0.5\%$ from subthreshold scaling fits, using the branch-and-bound decoder.
        For memory $\Lambda = p_L(d)/p_L(d+2)$, and for stability $\Lambda = P_{\rm fail}(r)/P_{\rm fail}(r+2)$.
    }
    \label{tab:lambda}
\end{table}

\cref{subsec:simulations:pure-leakage} confirmed the scalings of \cref{tab:min-weight} under a gate-only leakage noise model in the rotated surface code.
We repeat this comparison for a loss-only noise model appropriate to neutral atoms.
We first consider loss at CZ gates only, and then add handoff loss.
\Cref{subfig:loss_only_comparison:cz,subfig:loss_only_comparison:handoff} compare walking, moonwalking, and foliation on the rotated and unrotated codes, in $Z$-memory $(d,r)=(5,6)$ and $Z$-stability $(d,r)=(4,4)$ experiments with the branch-and-bound decoder.

\Cref{subfig:loss_only_comparison:cz} shows the results for CZ loss, without handoff loss.
The slopes agree with \cref{tab:min-weight}.
For memory, $\alpha \approx 1$ for moonwalking and foliation, and $\alpha \approx 0.8$ for rotated walking.
For stability, $\alpha \approx 1$ for foliation and $\alpha \approx 0.5$ for walking and moonwalking.
In walking and moonwalking circuits each atom lives for two rounds, so a single loss can bridge two time steps.
This leads to a reduction in the time-like error suppression compared to foliation.
We also observe a performance split between code geometries.
Without handoff loss, memory experiments favor the unrotated code, while stability experiments favor the rotated code.
We attribute this to entropic factors; the code with fewer minimum-weight failure paths has the lower error rate.
For memory, the unrotated code has a lower $p_L$ for every protocol because it has fewer minimum-weight logical operators.
For stability, the rotated code gives a lower $P_{\mathrm{fail}}$, because its distance-$4$ patch has fewer $Z$ stabilizers and thus fewer minimum-weight failure paths.

Next, we quantify the error suppression for each experiment.
The suppression factor is $\Lambda=p_L(d)/p_L(d+2)$ for memory and $\Lambda=P_{\mathrm{fail}}(r)/P_{\mathrm{fail}}(r+2)$ for stability.
We simulate memory with $d\in\{3,5\}$ and $r = 6$, and stability with $d=4$ and $r\in\{3,5,7\}$.
In both cases we set the CZ loss probability to $p_{\mathrm{CZ}} = 0.5\%$.
\Cref{tab:lambda} lists the computed suppression factors.
In memory, moonwalking and foliation give comparable space-like suppression.
In stability, foliation gives the largest time-like suppression on both code geometries.
Moonwalking outperforms walking in both experiments.
In memory, moonwalking also achieves the optimal space-like correction condition $t_{\rm E}\bnb=d$, which walking does not~\cite{pavlovich2026erasuresurfacecodecircuit}.
Furthermore, under CZ loss, an ancilla lost during a walking round skips its end-of-round role swap and remains on the same check into the next round, increasing the bulk CZ path multiplicity relative to moonwalking.

\Cref{subfig:loss_only_comparison:handoff} shows the results with handoff loss.
We set $p_{\mathrm{handoff}} = 2p_{\mathrm{CZ}}$.
Handoff loss penalizes the rotated SWAP-based circuits most, since they need the most handoffs per round; see~\cref{subsec:simulations:neutral-atom}.
In rotated memory, foliation is therefore far ahead of the SWAP-based circuits.
In unrotated memory, the inclusion of handoff loss narrows the lead of moonwalking.
Without handoff loss, moonwalking has a lower $p_L$ than foliation.
With handoff loss, the two are comparable.
In stability, the superior scaling of foliation outweighs its handoff overhead in both geometries.
Handoff loss also makes rotated walking worse than unrotated walking in stability, reversing the order without handoff loss.

\subsubsection{Realistic neutral-atom noise}

So far we have considered a loss-only noise model.
We now add additional sources of error realistic for a neutral-atom processor~\cite{bluvstein2025afaulttolerant}.
We describe our noise model in~\cref{app:harvard_noise_model}.
Besides CZ and handoff loss, the model includes anisotropic two-qubit Pauli errors of the CZ gate, Hadamard errors, and state-preparation and measurement (SPAM) errors.
All atoms that idle during a CZ layer suffer Pauli errors, and inter-zone shuttling and intra-zone rearrangement add decoherence.
Idle atoms, SPAM, shuttling, and rearrangement also cause loss.

We simulate $Z$-memory with $(d,r)=(7,6)$, and $Z$-stability with $(d,r)=(4,4)$ using the branch-and-bound decoder.
We define a parameter $\lambda\in[0,1]$ that shifts the infidelity of every operation from Pauli errors ($\lambda=0$) to loss ($\lambda=1$) at fixed state-averaged infidelity $p$.
At $\lambda=0.5$ every channel with both Pauli and loss components takes its nominal rates.

\Cref{fig:realistic_noise_interpolation} shows the results.
Increasing $\lambda$ moves all curves to larger $p$ and steepens the subthreshold scaling of memory from $\alpha \approx 0.5$ towards the erasure slope $\alpha \approx 1$.
In rotated memory, foliation outperforms walking and moonwalking for all $\lambda$.
The SWAP-based circuits need 8 CZ layers and about 16--20 handoffs per data qubit per round; see \cref{subsec:simulations:neutral-atom}.
In unrotated memory, moonwalking achieves lower error rates than foliation for $\lambda < 1$.
Foliation needs more CZ layers and handoffs per round than moonwalking, which add shuttling decoherence, spectator errors, and handoff loss.
Its shorter atom lifetime pays off as $\lambda$ grows; relative to unrotated walking, foliation improves with $\lambda$ and overtakes it at pure loss.
In stability, foliation outperforms walking and moonwalking in both geometries for all $\lambda$.
For $\lambda=1$, walking and moonwalking scale as $\alpha \approx 0.5$ and foliation as $\alpha \approx 1$, which gives foliation a large advantage.

\section{Conclusions}\label{sec:conclusions}

We have shown that the foliated surface code can optimally correct skip-gate leakage errors with only end-of-the-line leakage detection.
That is, with the branch-and-bound decoder, the number of correctable errors in memory, stability, and tCNOT experiments is equal to the number of detectable errors.
Our key insight was that optimal correction with end-of-the-line leakage detection requires an error correction circuit that avoids weight-2 hook errors.
We show that in SWAP-based circuits, the number of correctable time-like errors is only a fraction of the detectable errors.

From the neutral-atom specific simulations we find that the best circuit depends on both the experiment and the code geometry.
Foliation outperforms SWAP-based circuits in stability for both code geometries and for a wide range of parameters.
In memory however, the unrotated moonwalking circuit has the best performance when loss is small compared to computational errors.
The rotated SWAP-based circuits are penalized mostly by moving the $X$ and $Z$ ancilla qubits separately with a single AOD, so a second AOD would change this comparison.
Nonetheless, even with two AODs we would still expect foliation to give a benefit because of better subthreshold error scaling.

An important consequence of our work is to show that it is possible to trade off repeated application of costly mid-circuit erasure checks with comparatively easier single state-selective readout and two-qubit gates under the skip-gate leakage model.
Furthermore, our neutral-atom specific simulations motivate a need for co-designing the error correction circuit and code with the target logical operation.
Logical circuits where time-like errors are expected to dominate should use foliation while SWAP-based error correction should be used when space-like errors are expected to dominate.
  
A natural extension to our work is to investigate the performance of other codes like topological Floquet codes and quantum low-density parity check (qLDPC) codes with SWAP-based and foliated error correction.
It is straightforward to construct the foliation circuit for any CSS code and a naive implementation would double the qubit cost.
The nice structure of the surface code requiring roughly equal number of data and ancilla qubits made zero-overhead SWAP circuits possible.
Zero-overhead SWAP circuits have been found for Floquet codes as well~\cite{westerheim2026walking}.
However, due to a lack of symmetry, zero-overhead SWAP circuits are not known for general qLDPC codes.
Thus, like the foliated circuit, qLDPC SWAP circuits will also have additional overhead costs.
Moreover, there will be new challenges in decoding with SWAP and foliated circuits requiring development of new types of decoders.

Finally, a fruitful direction of research would be to find ways to optimally correct another common type of leakage effect: \textit{dephase-skip-gate leakage}. 
In this noise model, a qubit leaks during a CZ gate and causes a dephasing error on its partner qubit, and then skip subsequent gates.
Floquet codes may be tolerant of such errors~\cite{westerheim2026walking}, and thus studying SWAP-based and foliated Floquet code circuits would help design optimal approaches to correct for this kind of leakage, and thus develop more efficient erasure qubit architecture without mid-circuit erasure checks.

More broadly, our work motivates a general formulation of the principles that determine optimal error correction circuits for leakage errors.

\section*{Acknowledgments}

We thank Noah Shutty, Kaavya Sahay, Katie Chang, and Hanna Westerheim for useful discussions. 
This material is based upon work supported by the
U.S. Army Research Office under grant number W911NF-23-1-0051 and
U.S. Department of Energy, Office of Science, National Quantum Information Science Research Centers, Codesign Center for Quantum Advantage (C2QA) under contract number DE-SC0012704. The views and conclusions contained in this document are those of the authors and
should not be interpreted as representing the official policies, either expressed or implied, of the Army Research
Office or the U.S. Government. The U.S. Government is authorized to reproduce and distribute reprints for
Government purposes notwithstanding any copyright notation herein.

\textit{Data availability---} Code and data for the gate-only leakage simulations are provided at Ref.~\cite{bi2026foliatedcode}, which builds on Ref.~\cite{pavlovich2026leakagecode}.
Data for the neutral-atom noise simulations is provided at Ref.~\cite{neutralatomdata}.

\bibliography{refs}

\appendix

\crefalias{section}{appendix}
\crefalias{subsection}{subappendix}

\setcounter{equation}{0}

\section{Circuits and data}\label{app:circuits}

\begin{figure*}
    \centering
    \includegraphics[width=\textwidth,height=0.86\textheight,keepaspectratio]{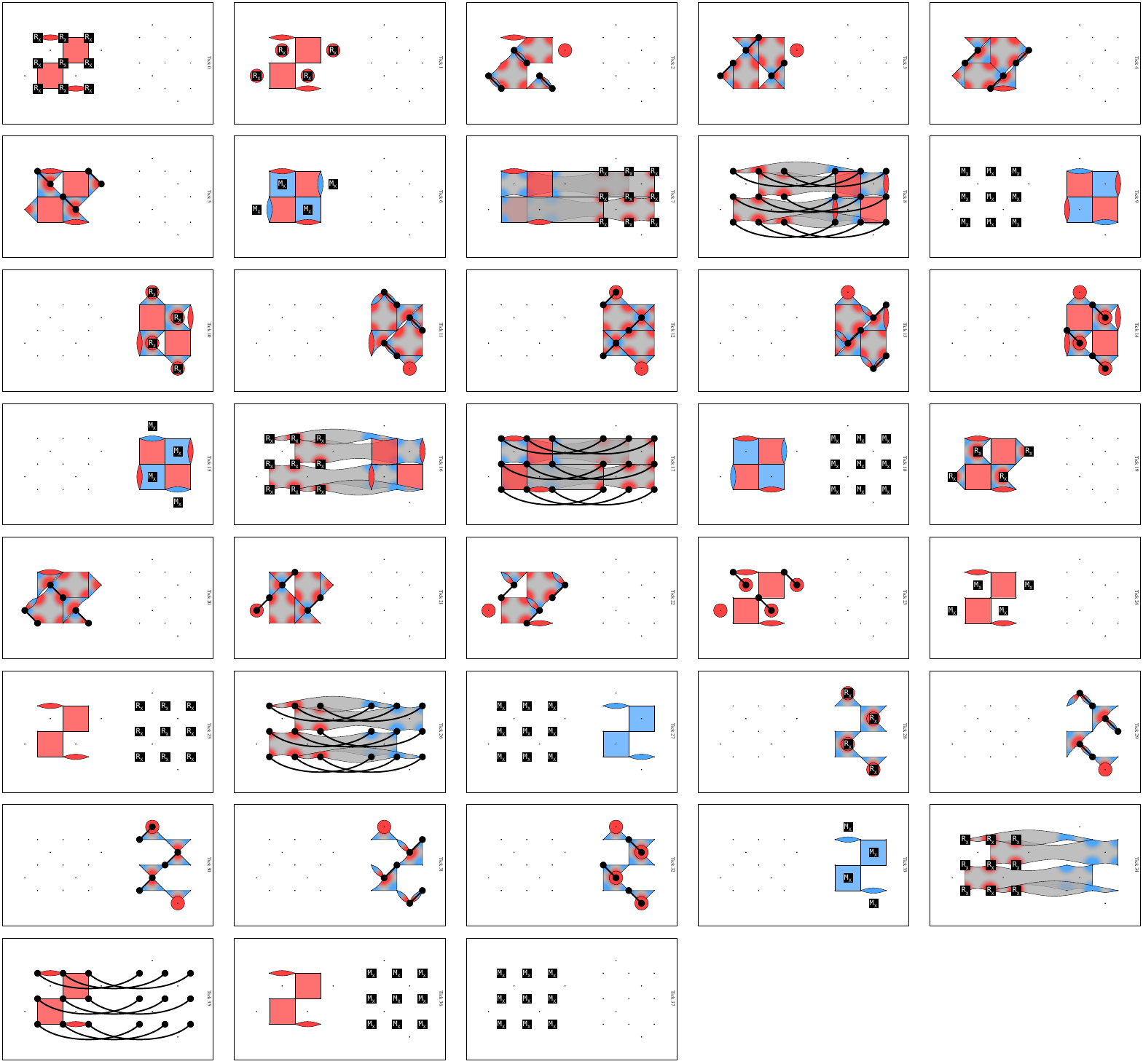}
    \caption{
        {\bf Basic foliation circuit, memory.}
        Stim detector slices for the circuit described in~\cref{subsec:foliation:circuit}.
        Here we show the circuit for an $X$-memory experiment of the distance-$3$ rotated surface code with two rounds of stabilizer measurements.
        The $Z$ and $X$ stabilizer measurements occur sequentially, separated by a teleportation layer.
    }
    \label{fig:circuit-basic}
\end{figure*}

\begin{figure*}
    \centering
    \includegraphics[width=\textwidth]{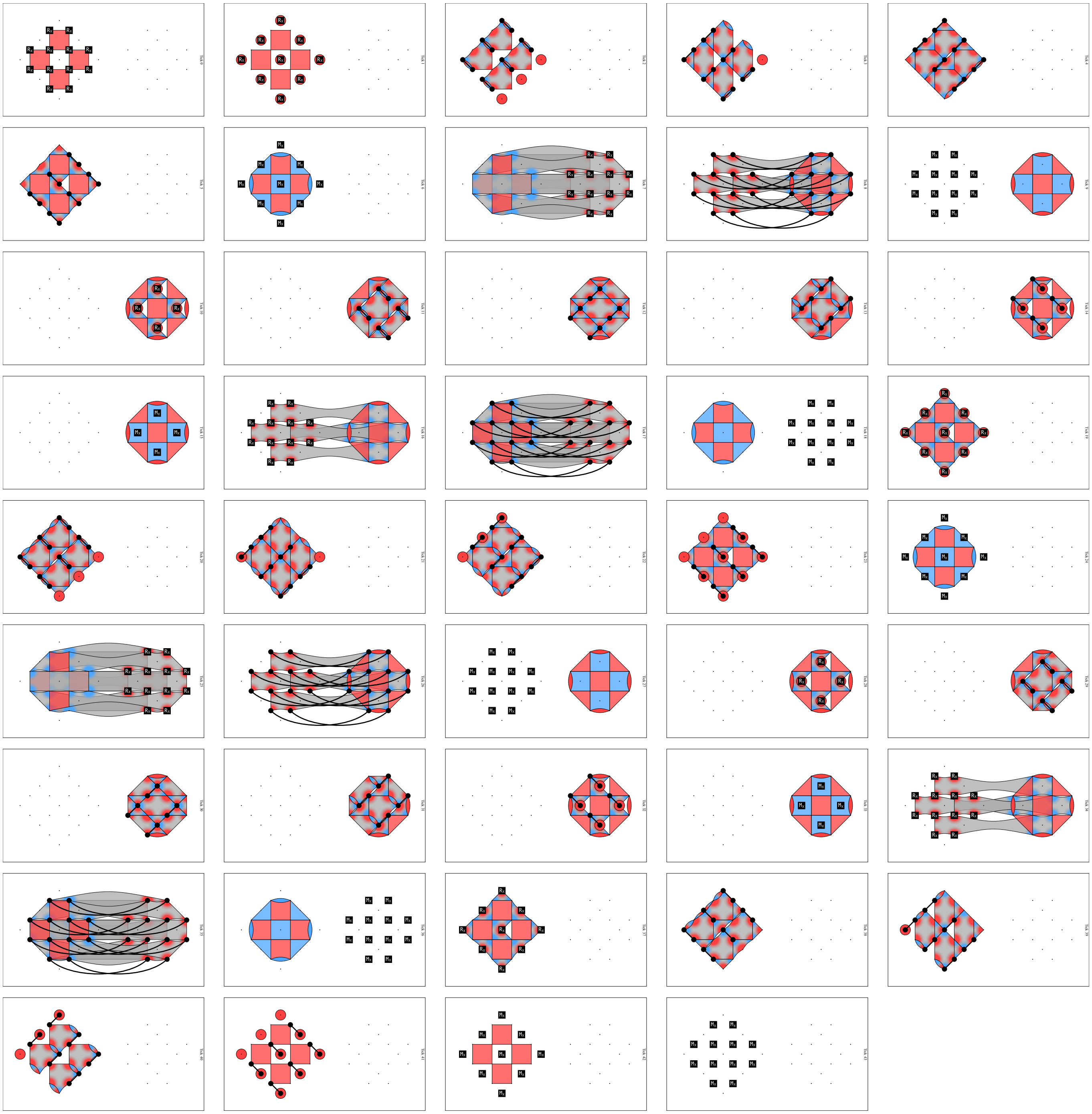}
    \caption{
        {\bf Basic foliation circuit, stability.}
        Stim detector slices for the circuit described in~\cref{subsec:foliation:circuit}.
        Here we show the circuit for a $Z$-stability experiment with a distance-$4$ patch and $r=3$ $Z$-stabilizer measurements.
    }
    \label{fig:circuit-basic-stability}
\end{figure*}

\begin{figure*}
    \centering
    \includegraphics[width=\textwidth]{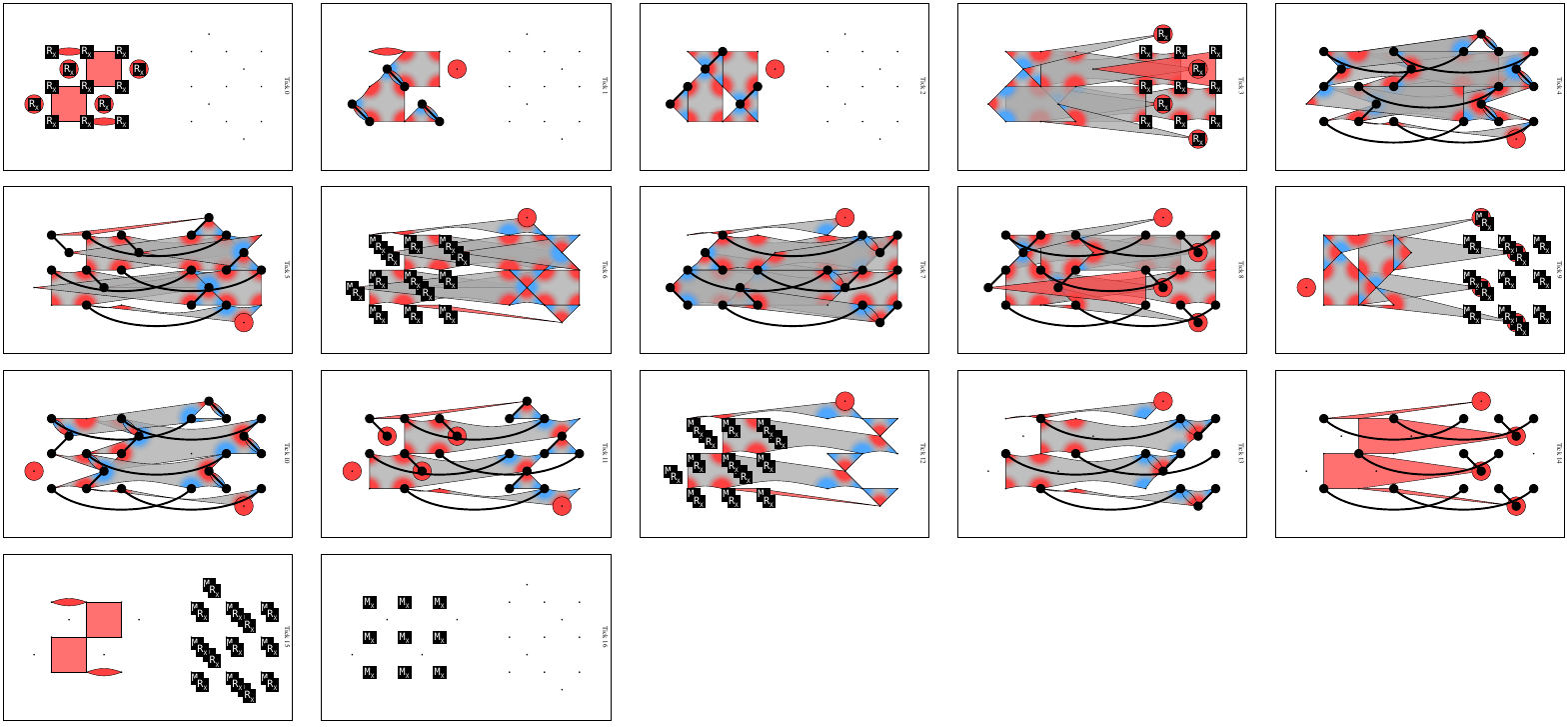}
    \caption{
        {\bf Compact foliation circuit, memory.}
        Stim detector slices for the circuit described in~\cref{subsec:foliation:compact}.
        Here we show the circuit for an $X$-memory experiment of the distance-$3$ rotated surface code with two rounds of stabilizer measurements.
        Each layer of the compact circuit interleaves $Z$ and $X$ stabilizer measurements, reducing the circuit depth.
    }
    \label{fig:circuit-compact}
\end{figure*}

\begin{figure*}
    \centering
    \includegraphics[width=\textwidth]{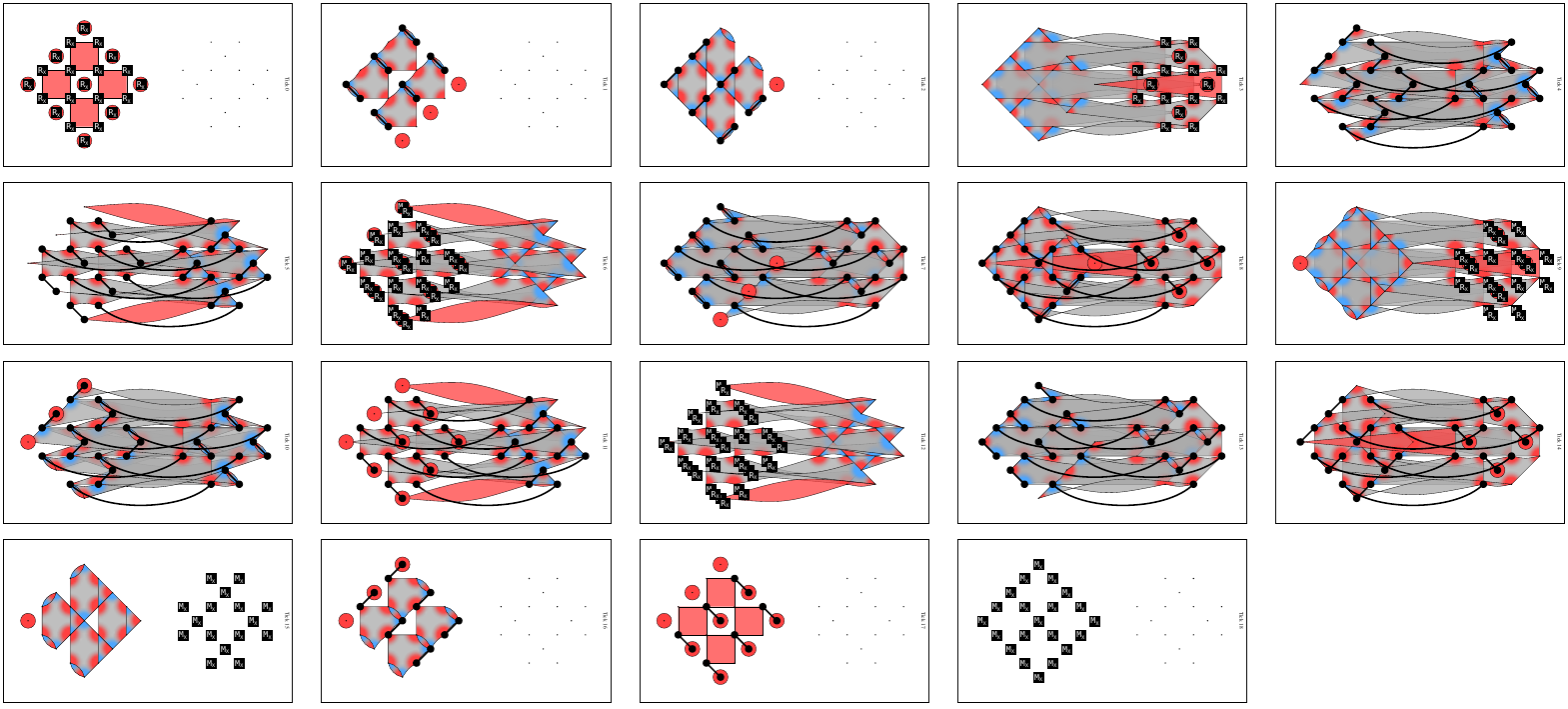}
    \caption{
        {\bf Compact foliation circuit, stability.}
        Stim detector slices for the circuit described in~\cref{subsec:foliation:compact}.
        Here we show the circuit for a $Z$-stability experiment with a distance-$4$ patch and $r=3$ $Z$-stabilizer measurements.
    }
    \label{fig:circuit-compact-stability}
\end{figure*}

\begin{figure*}
    \centering
    \includegraphics[width=\textwidth]{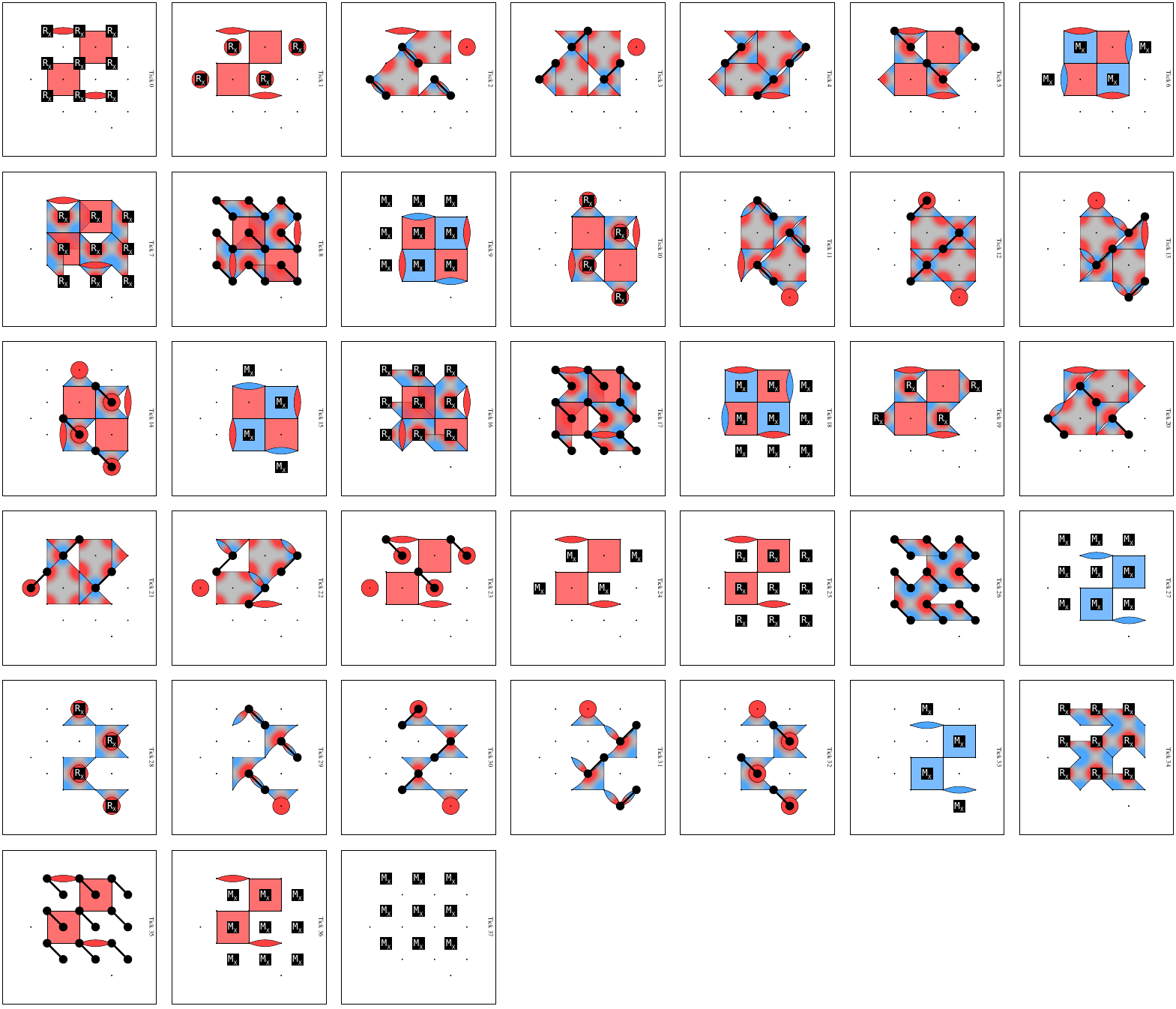}
    \caption{
        {\bf Planar foliation circuit, memory.}
        Stim detector slices with $d=3$ and two rounds, using the same schedule as~\cref{fig:circuit-basic}.
        The second lattice is displaced by one site, so that the two lattices overlap on a single plane and all gates are local.
    }
    \label{fig:circuit-planar}
\end{figure*}

\Cref{fig:circuit-basic,fig:circuit-basic-stability,fig:circuit-compact,fig:circuit-compact-stability,fig:circuit-planar} show Stim detector slices for the basic, compact and planar foliation circuits.
Memory circuits are for a distance-3 $X$-memory experiment with two complete stabilizer measurement rounds (i.e. both $X$ and $Z$ stabilizers), and stability circuits are for a distance-4 $Z$-stability experiment with $r=3$.
Red regions represent $X$ detectors, which flip in the presence of $Z$ errors, and blue regions represent $Z$ detectors, which flip in the presence of $X$ errors.

\begin{figure*}
    \centering
    \includegraphics[width=\textwidth]{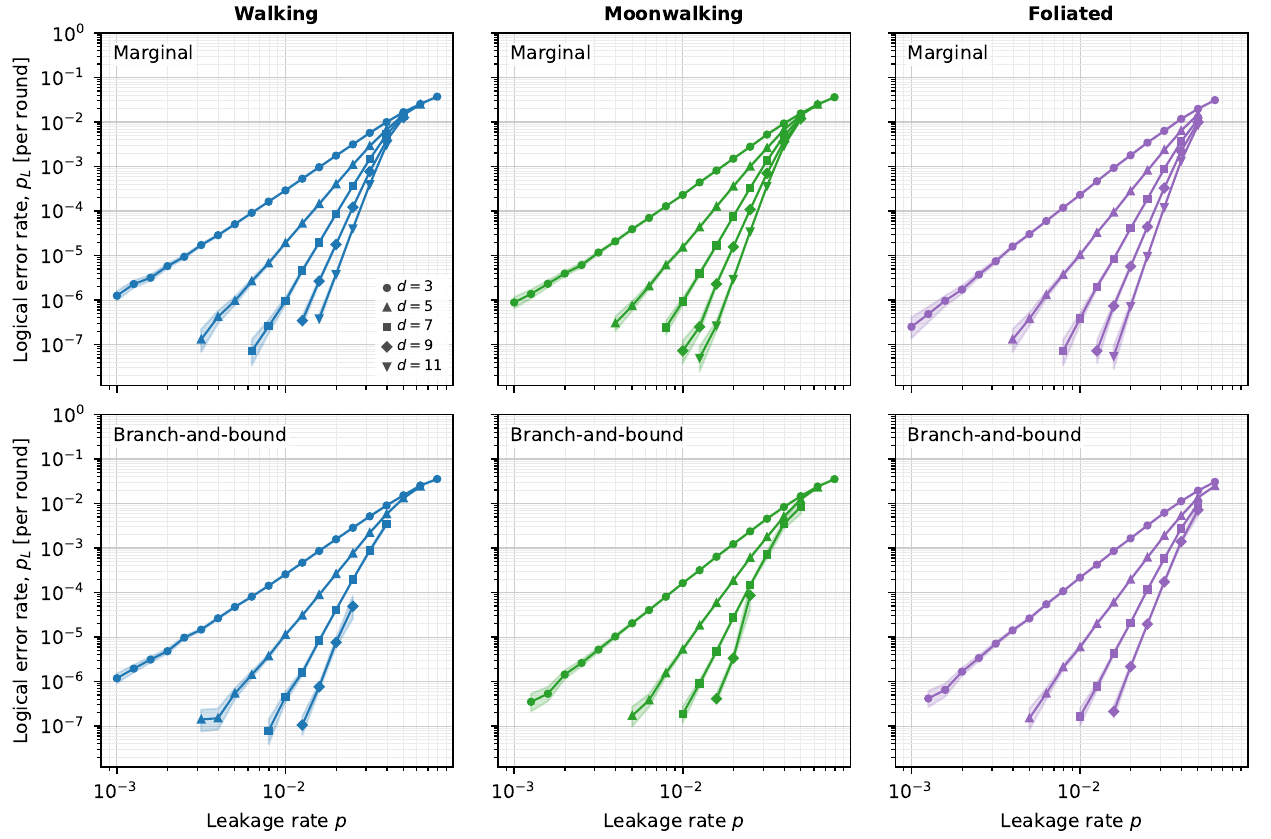}
    \caption{
        {\bf Gate-only leakage, memory.}
        Logical error rate per round $p_L$ against leakage rate $p$, with rounds $r = 4d$.
        The fits of~\cref{fig:subthreshold} use $d \in \{5, 7, 9\}$.
        Bands show $96\%$ confidence intervals.
    }
    \label{app-fig:memory-ler}
\end{figure*}

\begin{figure*}
    \centering
    \includegraphics[width=\textwidth]{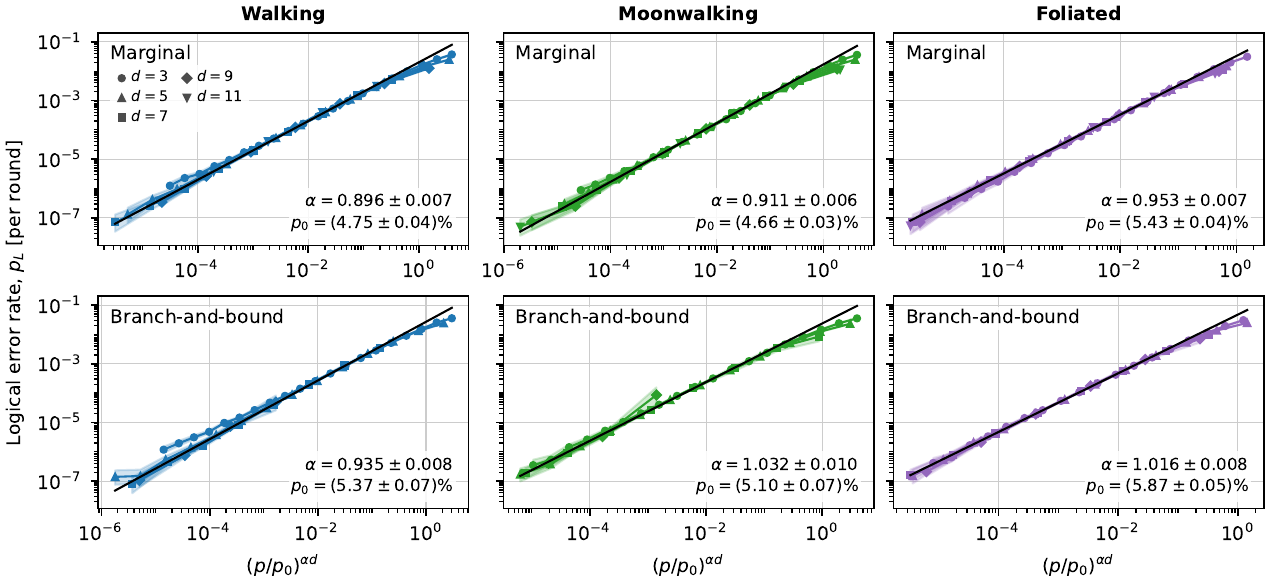}
    \caption{{\bf Subthreshold scaling collapse, memory.}
        Data of~\cref{app-fig:memory-ler} rescaled by the fitted $\alpha$ and $p_0$ of~\cref{fig:subthreshold}.
    }
    \label{app-fig:memory-subth}
\end{figure*}

\begin{figure*}
    \centering
    \includegraphics[width=\textwidth]{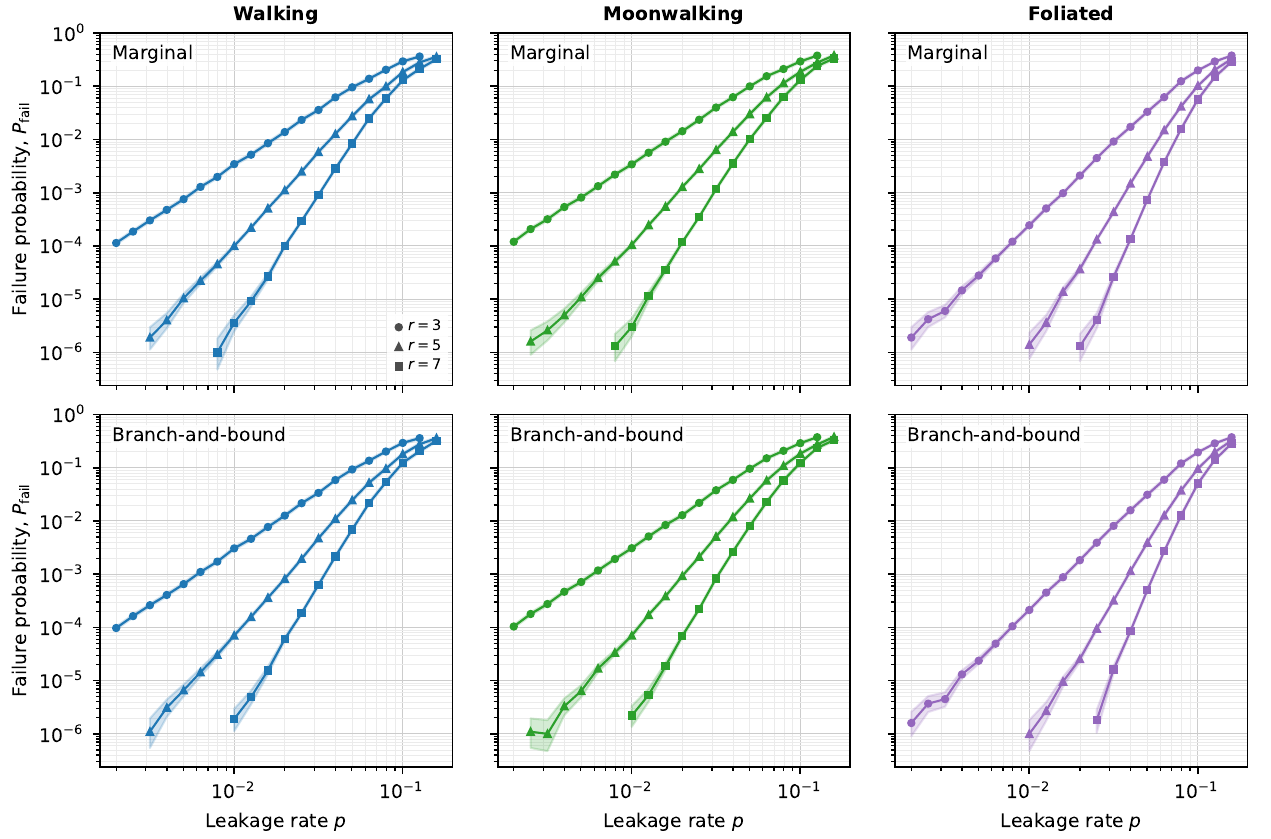}
    \caption{
        {\bf Gate-only leakage, stability.}
        Failure probability per shot $P_{\rm fail}$ against leakage rate $p$, for distance $d = 4$.
        Bands show $96\%$ confidence intervals.
    }
    \label{app-fig:stability-ler}
\end{figure*}

\begin{figure*}
    \centering
    \includegraphics[width=\textwidth]{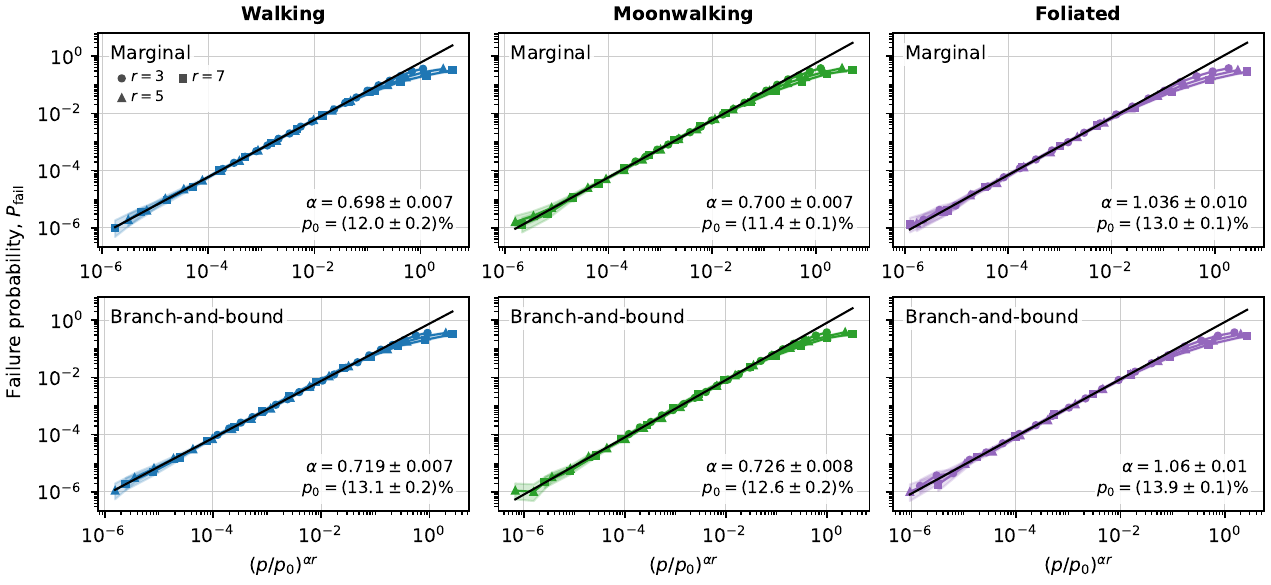}
    \caption{
        {\bf Subthreshold scaling collapse, stability.}
        Data of~\cref{app-fig:stability-ler} rescaled by the fitted $\alpha$ and $p_0$ of~\cref{fig:subthreshold}.
    }
    \label{app-fig:stability-subth}
\end{figure*}

\Cref{app-fig:memory-ler,app-fig:memory-subth,app-fig:stability-ler,app-fig:stability-subth} show the $X$-memory logical error rates and $Z$-stability failure probabilities behind the subthreshold fits (\cref{fig:subthreshold}), and their collapse onto the fitted subthreshold scaling model.
Code and data are provided at Ref.~\cite{bi2026foliatedcode}.

\clearpage
\onecolumngrid %
\begin{figure*}[t!]
    \centering
    \includegraphics[width=1\textwidth]{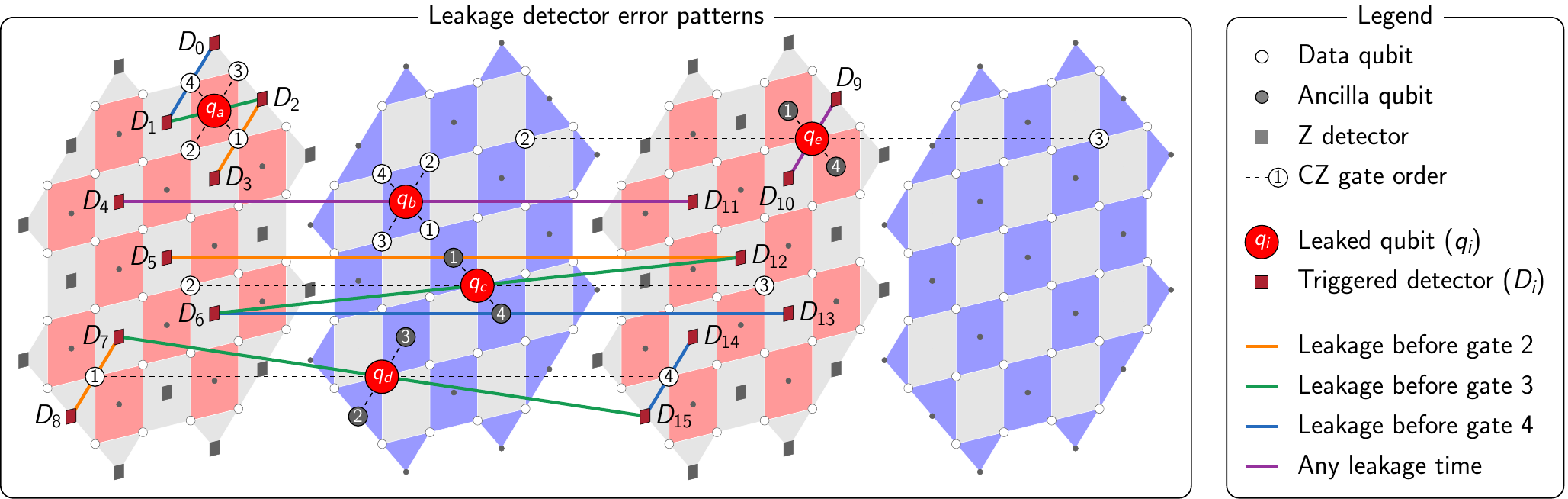}
    \caption{
        {\bf Leakage detector error patterns.} Shown for a two-round $Z$-stability experiment with the foliated circuit in the compact gate ordering.
        Time runs from left to right.
        Open circles, filled circles, and squares denote data qubits, ancilla qubits, and $Z$ detectors, respectively; triggered detectors $D_i$ are dark red.
        Red and blue plaquettes indicate active $X$ and $Z$ checks; gray plaquettes indicate inactive checks.
        Red circles mark representative leaked qubits: $X$ ancilla $q_a$, $Z$ ancilla $q_b$, and data qubits in the $Z$-check layer ($q_c$, $q_d$) or the $X$-check layer ($q_e$).
        Circles numbered 1--4 mark the CZ partners of each leaked qubit in gate order; dashed lines connect partners in other layers.
        Solid orange, green, and blue edges show detector pairs that can be triggered by leakage immediately before gates 2, 3, and 4, respectively.
        A single purple edge indicates that the leaked qubit produces the same edge regardless of when the leakage occurs.
    }
    \label{fig:dem-graph}
\end{figure*}
\twocolumngrid

\section{Memory and stability scaling}\label{app:scaling}

In this appendix, we derive the minimum weights of undetectable and uncorrectable leakage errors in memory and stability experiments, summarized in \cref{tab:min-weight}.
In \cref{app:scaling:patterns} we map single leakage errors in the foliated circuit to detector error patterns.
In \cref{app:scaling:undetectable} we use these patterns to derive $d_{\rm L}$, and in \cref{app:mem-stab-scaling:min-weight} we derive $t_{\rm E}$ for the branch-and-bound and marginal decoders, including the connection to weight-2 hook errors.
In \cref{app:walking-stability} we apply the same analysis to walking and moonwalking stability experiments.

\subsection{Leakage detector error patterns}\label{app:scaling:patterns}

Our argument is based on the decoding graph, so we will start by going over the detector error patterns that a single-qubit leakage error will cause.

\Cref{fig:dem-graph} shows a distance-5, two-round stability experiment implemented with a compact foliation circuit.
The data qubits are initialized in the $X$ basis in the first layer, teleported through the circuit, and measured in the $X$ basis in the final layer.
The logical observable is the product of all $Z$ checks in the final layer.
Therefore, only $Z$ detectors need to be considered.
We retain the conventional N/Z-shaped syndrome extraction gate orderings of the rotated surface code.

There are four types of qubits in this circuit: $X$ ancillas ($q_a$), $Z$ ancillas ($q_b$), data qubits in the $Z$-check layer ($q_c$ and $q_d$), and data qubits in the $X$-check layer ($q_e$).
Leakage of each qubit type can produce different detector error patterns.
For example, in \cref{fig:dem-graph} the qubit $q_a$ can leak anywhere in its lifetime.
If $q_a$ leaks before gate 1, no detector is triggered.
If it leaks before gate 2, detectors $D_2$ and $D_3$ may be triggered, corresponding to the orange edge connecting $D_2$ and $D_3$ in the decoding graph.
The green and blue edges similarly correspond to the leakage of $q_a$ before gates 3 and 4, respectively.
For a given leaked qubit, we call an edge in the decoding graph a \emph{candidate edge} if leakage of that qubit can produce the edge for at least one leakage time.
Leakage of $q_c$ or $q_d$ can also produce three distinct detector error patterns that follow the same color convention.
Under the compact gate ordering, $q_c$ and $q_d$ experience different gate sequences and therefore produce different error patterns.
In the basic circuit, however, leakage of a data qubit in the $Z$ layer produces only the error patterns associated with the leakage of $q_c$.
A single purple edge indicates that the leaked qubit produces the same detector error pattern regardless of when the leakage occurs.

Although the detailed detector pattern depends on the qubit type and the time of leakage, all of the cases in \cref{fig:dem-graph} share the property relevant to the fault-distance argument: a single leakage error can contribute at most one elementary edge to the support of a shortest logical string.
This contribution can be either space-like or time-like.
We use this one-edge contribution property below to relate the local detector patterns to the minimum-weight undetectable and uncorrectable errors.

\subsection{Minimum-weight undetectable errors}\label{app:scaling:undetectable}

An undetectable logical error is represented in the decoding graph by a nontrivial logical chain with no detector endpoints.
\Cref{fig:single-memory} shows a single $X$-check layer of the foliated memory circuit.
The marked $X$ operators form part of the logical $Z_L$ observable, while the marked $Z$ operators form a shortest $X$-type logical error connecting the top and bottom space boundaries.

Each leakage error contributes at most one elementary edge along a shortest logical string, so at least $d$ leakage errors are required.
Notice that the $D_1$--$D_2$ edge in \cref{fig:dem-graph} may appear equivalent to a weight-2 edge.
However, it is not directed along a shortest logical string and therefore cannot reduce $d_{\rm L}$.
This property results from the correct gate ordering of each check in the circuit.
As a result, at least $d$ leakage errors are required to form an undetectable logical error.
This bound is attained by placing the leakage errors along such a minimum-length string.
Therefore, for a memory experiment
\begin{equation}
    d_{\rm L} = d \qquad \text{(memory)}.
\end{equation}

For a stability experiment, the corresponding logical chain is time-like and connects the two open time boundaries of the decoding graph.
Its minimum length is the number of rounds $r$.
Since each leakage error contributes at most one elementary edge along this chain, at least $r$ leakage errors are required.
A straight time-like string attains this bound.
Hence, for a stability experiment
\begin{equation}
    d_{\rm L} = r \qquad \text{(stability)}.
\end{equation}

\subsection{Minimum-weight uncorrectable errors}\label{app:mem-stab-scaling:min-weight}

\subsubsection{Branch-and-bound decoder}

Recall that state-selective readout identifies the leaked qubit but not its leakage time between initialization and measurement.
Because the qubit can leak only once in this interval, the candidate leakage times are mutually exclusive and therefore form a disjoint probability distribution~\cite{pavlovich2026erasuresurfacecodecircuit}.
The branch-and-bound decoder enforces this disjointness constraint, so its correction is consistent with exactly one leakage-time branch for each detected leakage error.
Let $E$ denote the effective Pauli error associated with the actual leakage error and $C$ the correction selected by the decoder.
A decoding failure occurs when the residual mismatch $C\oplus E$ forms a nontrivial logical operator.

For each leakage error, the residual mismatch $C\oplus E$ contributes at most one weight-1 edge to the decoding graph, either space-like or time-like.
For example, if $E$ corresponds to the edge $D_5$--$D_{12}$ in \cref{fig:dem-graph} and the correction $C$ corresponds to the edge $D_6$--$D_{12}$, then the mismatch $C\oplus E$ turns out to be $D_5$--$D_6$, which is a weight-1 space-like edge in the decoding graph.
Therefore, a logical failure requires at least $d$ leakage errors in a memory experiment or $r$ leakage errors in a stability experiment.
Consequently, for the branch-and-bound decoder,
\begin{equation}
    \begin{aligned}
        t_{\rm E}\bnb &= d_{\rm L} = d \qquad \text{(memory)},\\
        t_{\rm E}\bnb &= d_{\rm L} = r \qquad \text{(stability)}.
    \end{aligned}
\end{equation}
The branch-and-bound decoder therefore saturates the fault distance of the foliation circuit in both experiments.

\subsubsection{Connection to weight-2 hook errors}

The residual mismatch $C\oplus E$ in the decoding graph provides an equivalent representation of the hook error configurations analyzed in \cref{sec:foliation}.
In each configuration, we identify the first $X$ error with the effective Pauli error $E$ induced by leakage of the data qubit, and the second with the correction $C$ selected by the decoder.
\cref{tab:hook-error-mapping} summarizes this correspondence for leakage of $q_a$, $q_c$ and $q_d$.

We write $e_{a,b}$ for the edge joining detectors $D_a$ and $D_b$.
The mismatches in Eqs.~\eqref{eq:anc-cz1}--\eqref{eq:data-cz1} and~\eqref{eq:data-cz3} correspond to a weight-1 space-like edge.
In \cref{eq:data-cz2-tele,eq:compact-mixed}, the mismatches lead to one space-like and one time-like error.
In \cref{eq:data-cz2-check,eq:compact-both}, the two mismatches both give pairs of time-like edges.
Each edge has weight 1 and lies on a distinct logical string.
Thus no configuration forms a weight-2 edge along a single logical string.
This corresponds exactly to the preceding analysis showing the absence of weight-2 hook errors.

\begin{table}
    \centering
    \small
    \renewcommand{\arraystretch}{1.4}
    \setlength{\tabcolsep}{4pt}
    \begin{tabular*}{\columnwidth}{@{\extracolsep{\fill}}lccl}
        \toprule
        Case & $E$ & $C$ & $C\oplus E$ \\
        \midrule
        \cref{eq:anc-cz1} & $\varnothing$ & $e_{2,3}$ & $e_{2,3}$ \\
        \cref{eq:anc-cz2v} & $\varnothing$ & $e_{1,2}$ & $e_{1,2}$ \\
        \cref{eq:anc-cz3} & $\varnothing$ & $e_{0,1}$ & $e_{0,1}$ \\
        \midrule
        \cref{eq:data-cz1} & $\varnothing$ & $e_{7,8}$ & $e_{7,8}$ \\
        \cref{eq:data-cz2-tele} & $\varnothing$ & $e_{7,15}$ & $e_{7,8}\oplus e_{8,15}$\\
        \cref{eq:data-cz2-check} & $e_{7,8}$ & $e_{14,15}$ & $e_{7,14}\oplus e_{8,15}$\\
        \cref{eq:data-cz3} & $\varnothing$ & $e_{14,15}$ & $e_{14,15}$\\
        \midrule
        \cref{eq:compact-mixed} & $\varnothing$ & $e_{6,12}$ & $e_{5,6}\oplus e_{5,12}$\\
        \cref{eq:compact-both} & $e_{5,12}$ & $e_{6,13}$ & $e_{5,12}\oplus e_{6,13}$\\
        \bottomrule
    \end{tabular*}

    \caption{{\bf Hook errors in the decoding graph.} Correspondence between the hook-error configurations of \cref{sec:foliation} and the mismatches $C\oplus E$ in the decoding graph.}
    \label{tab:hook-error-mapping}
\end{table}

\subsubsection{Marginal decoder}
\label{app:marginal-te}

The marginal ({\tt marg}) decoder merges the candidate leakage times into one matching graph and decodes it by minimum-weight perfect matching (MWPM).
Since the disjointness constraint is discarded, it may select multiple candidate edges generated by a single leakage error.
Here, we show that, under gate-only leakage noise, $t_{\rm E}\marg\sim 2d/3$ for memory experiments and $t_{\rm E}\marg\sim r$ for stability experiments.

We first determine the weights of the colored edges in \cref{fig:dem-graph}.
Consider $q_a$, which may leak before any of gates 1--4.
We assign each location a probability of approximately $1/4$.
For leakage before gate 2, we construct the Pauli envelope by applying a depolarizing channel at that location.
A $Z$ or $Y$ error, each with probability $1/4$, flips detectors $D_2$ and $D_3$, producing the orange edge in the decoding graph.
The probability associated with this edge is thus $p=2(1/4)(1/4)=1/8$, which gives the matching weight $w_0=\log((1-p)/p)=\log 7$.

Applying the same calculation to the remaining colored edges, we find that $D_0$--$D_1$, $D_1$--$D_2$, $D_5$--$D_{12}$, $D_6$--$D_{12}$, $D_6$--$D_{13}$, $D_7$--$D_8$, $D_7$--$D_{15}$, and $D_{14}$--$D_{15}$ all have weight $w_0$.
The edges $D_4$--$D_{11}$ and $D_9$--$D_{10}$ instead have weight $w_1=\log\frac{1-1/2}{1/2}=0$.

We next determine the minimum number of leakage errors required to cause the marginal decoder to fail.
A failure configuration must satisfy two conditions:
\begin{enumerate}

    \item 
    \textit{Connectivity condition.}
    The candidate edges for the leaked qubits must form a continuous path connecting the two open space boundaries.

    \item 
    \textit{Weight condition.}
    The correction $C$ returned by MWPM must have weight no greater than that of the true error chain $E$ generated by the leakage errors.
    Namely, $w(C)\leq w(E)$.
    
\end{enumerate}
For each leakage error $i$, let $S_i$ denote the set of candidate edges for the qubit that leaked.
Leakage errors $i$ and $i'$ belong to the same \emph{elementary block} whenever their candidate edge sets have a nonempty intersection, $S_i\cap S_{i'}\neq\varnothing$.
Let $B_j$ denote the set of error indices in the $j$th elementary block.
Distinct elementary blocks therefore have disjoint edge supports, although they may share detector vertices.

\textit{Lemma.---} For the $j$th elementary block in the bulk of the decoding graph, with $k_j = |B_j|$ leakage errors, let $U_j = \bigcup_{i\in B_j} S_i$ denote its candidate edge set.
Define $C_j = C \cap U_j$ and $E_j = E \cap U_j$.
Let $\ell_j$ be the effective spatial advance of $C_j \oplus E_j$, measured in elementary steps along the logical direction connecting the two open space boundaries.
Define
\begin{equation}
    \Delta_j=w(C_j)-w(E_j).
\end{equation}
Then the block satisfies the weight--advance bound
\begin{equation}
    \ell_j\leq \frac{3k_j}{2}+\frac{\Delta_j}{2w_0}.
    \label{eq:local-weight-advance}
\end{equation}

\textit{Proof.---} For a shared edge $e \in S_i \cap S_{i'}$ with $i \ne i'$, merging the independent probability contributions from the two leaked qubits preserves the bound $0 \leq w(e) \leq w_0$.
Each physical leakage error contributes at most one edge to $E$.
Thus $w(E_j)\leq k_jw_0$ and $\Delta_j\geq-k_jw_0$, so \cref{eq:local-weight-advance} follows immediately whenever $\ell_j\leq k_j$.
Suppose the $j$th elementary block contains a candidate edge set whose pattern differs from that of $q_a$.
As shown in \cref{fig:dem-graph}, this set spans at most one elementary step along the logical direction.
Starting with this set, add the remaining $k_j-1$ candidate edge sets one at a time.
Each set needs to share at least one edge with the current union.
The geometry in \cref{fig:dem-graph} ensures that each addition increases the spatial span of the union by at most one elementary step.
Thus the full union $U_j$ spans at most $1+(k_j-1)=k_j$ elementary steps.
Since $C_j\oplus E_j\subseteq U_j$, the effective spatial advance of this elementary block satisfies $\ell_j\leq k_j$, so naturally satisfies \cref{eq:local-weight-advance}.

Now consider the case in which every candidate edge set in the $j$th elementary block has the pattern shown for $q_a$ in \cref{fig:dem-graph}.
Then $U_j$ forms a chain spanning $k_j+1$ elementary steps along the logical direction.
The preceding argument only covers $\ell_j\leq k_j$, so it remains to consider $\ell_j=k_j+1$.
To achieve this advance, $C_j\oplus E_j$ must contain all $k_j+2$ unshared edges of the chain, each of weight $w_0$.
Since $E_j$ contains at most $k_j$ edges, at least two of these edges must belong to $C_j$, giving $w(C_j)\geq 2w_0$.
Together with $w(E_j)\leq k_jw_0$, this yields $\Delta_j\geq(2-k_j)w_0$ and satisfies \cref{eq:local-weight-advance}.
This argument applies for all $k_j\geq 1$.

\begin{figure}
    \centering
    \includegraphics[width=0.65\columnwidth]{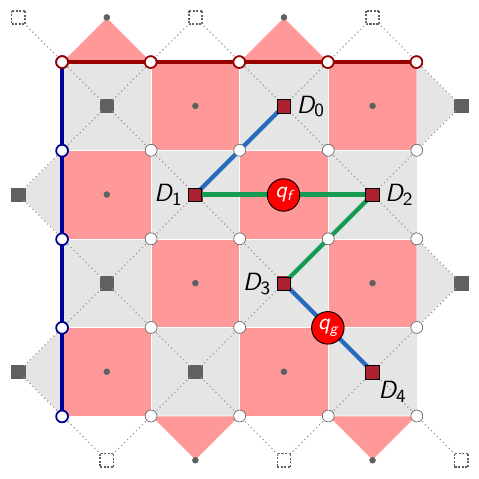}
    \caption{
        {\bf Saturating marginal-decoder configuration.} One $X$-check layer of a distance-5 foliated $Z$-memory circuit.
        The red $X$ string forms part of the logical $Z_L$ observable, while the blue $Z$ string marks a shortest $X$-type logical operator.
        Red circles mark the leaked ancilla $q_f$ and the leaked data qubit $q_g$ at the same spatial location in the next $Z$-check layer.
        Leakage of $q_f$ generates three candidate edges, $D_0$--$D_1$, $D_1$--$D_2$, and $D_2$--$D_3$.
        Leakage of $q_g$ generates another set of three candidate edges, including $D_3$--$D_4$.
        The other two are not shown here.
        Consider the case in which the true error $E_j$ corresponds to the blue edges $D_0$--$D_1$ and $D_3$--$D_4$, while the marginal decoder selects the green edges $D_1$--$D_2$ and $D_2$--$D_3$ as the correction $C_j$.
        With $w(C_j)=w(E_j)=2w_0$, the mismatch $C\oplus E$ advances three elementary spatial steps, giving $(k_j,\ell_j,\Delta_j)=(2,3,0)$ and saturating \cref{eq:local-weight-advance}.
    }
    \label{fig:single-memory}
\end{figure}

The bound is tight.
In \cref{fig:single-memory}, leakage of $q_f$ produces three candidate edges: $D_0$--$D_1$, $D_1$--$D_2$, and $D_2$--$D_3$.
Leakage of $q_g$, the data qubit at the same spatial location in the next $Z$-check layer, also produces three candidate edges, each of weight $w_0$.
These three edges overlap in the two-dimensional projection shown in the figure.
Let the two physical leakage errors contribute the blue edges $D_0$--$D_1$ and $D_3$--$D_4$, and let the decoder select the green edges $D_1$--$D_2$ and $D_2$--$D_3$.
The resulting block has
\begin{equation}
    (k_j,\ell_j,\Delta_j)=(2,3,0),
\end{equation}
and therefore saturates \cref{eq:local-weight-advance}.
\hfill$\square$

As immediate consequences, a single leakage error with $\Delta_j\leq0$ has $\ell_j\leq3/2$, and hence $\ell_j\leq1$ because $\ell_j$ is integer-valued.
Two leakage errors with $\Delta_j\leq0$ have $\ell_j\leq3$, with equality in the configuration above.

We now extend the local counting argument to a distance-$d$ memory experiment.
Let $k=\sum_jk_j$ be the total number of leakage errors and let $\ell$ be the net spatial advance.
We have $\ell\leq\sum_j\ell_j$, while the disjoint elementary blocks give $\sum_j\Delta_j=w(C)-w(E)$.
Summing \cref{eq:local-weight-advance} and using the weight condition gives
\begin{equation}
    \ell
    \leq \frac{3k}{2}+\frac{w(C)-w(E)}{2w_0}
    \leq \frac{3k}{2}.
    \label{eq:global-weight-advance}
\end{equation}
The bulk portion of a nontrivial logical mismatch must span at least $d-O(1)$ spatial steps, where the constant correction accounts for the boundary regions excluded from the local bound.
\Cref{eq:global-weight-advance} therefore implies $k\geq \frac{2d}{3}-O(1)$, and thus
\begin{equation}
    t_{\rm E}\marg\geq \frac{2d}{3}-O(1).
\end{equation}

Conversely, tiling a shortest logical path with the two-leakage, three-step pattern in \cref{fig:single-memory} and adjusting the configuration near the spatial boundaries produces a logical path satisfying the weight condition $w(C)\leq w(E)$.
The boundary adjustments require only $O(1)$ additional leakage errors, so the full failure configuration contains $2d/3+O(1)$ leakage errors.
This construction gives
\begin{equation}
    t_{\rm E}\marg\leq \frac{2d}{3}+O(1).
\end{equation}
Combining the lower and upper bounds gives
\begin{equation}
    t_{\rm E}\marg=\frac{2d}{3}+O(1)
    \sim \frac{2d}{3}
    \qquad \text{(memory)}.
\end{equation}

For the stability experiment, each leakage pattern in \cref{fig:dem-graph} contributes at most one time-like edge of the residual mismatch $C\oplus E$, regardless of the number of candidate edges selected by the marginal decoder.
A nontrivial logical path must connect the two open time boundaries and hence span $r$ time-like edges.
A decoding failure therefore requires at least $r$ leakage errors.
Together with $t_{\rm E}\marg\leq d_{\rm L}=r$,  the marginal decoder saturates the fault distance:
\begin{equation}
    t_{\rm E}\marg=d_{\rm L}=r
    \qquad \text{(stability)}.
\end{equation}

\subsection{SWAP-based stability experiments}
\label{app:walking-stability}

We apply the same fault distance analysis to the SWAP-based circuits considered in~\cite{pavlovich2026erasuresurfacecodecircuit}.
Without mid-circuit erasure checks, each bulk qubit undergoes eight gates between initialization and measurement, compared with four in the foliated circuit.
Unlike in the foliated circuit, which uses only CZ gates, each qubit in these circuits switches between control and target roles during its lifetime, so each leakage error can produce more complex detector error patterns.
For each potentially leaked qubit, we enumerate the detector error patterns for all possible leakage locations to obtain its candidate set.

As in the foliated circuit, each leakage error contributes at most one elementary time-like edge along a shortest temporal logical string.
An undetectable logical chain connecting the two open time boundaries therefore requires at least $r$ leakage errors.
Hence:
\begin{equation}
    d_{\rm L} = r
    \qquad \text{(stability)}.
    \label{eq:walking-stability-dl}
\end{equation}

The longer interval between initialization and measurement and the role switching between control and target enlarge the candidate sets.
The branch-and-bound decoder still selects at most one candidate edge for each detected leakage error, enforcing the mutual exclusivity of its leakage times.
However, enumerating the allowed pairs of physical error $E$ and correction $C$ shows that the residual mismatch $C\oplus E$ can advance a single temporal logical string by up to two elementary steps, corresponding to a weight-2 time-like edge.
Thus a logical failure requires at least $r/2-O(1)$ leakage errors.
Conversely, concatenating these weight-2 mismatch patterns along a temporal logical chain yields two allowed error realizations with the same detected leakage errors and detector syndrome but different logical outcomes, using $r/2+O(1)$ leakage errors.
The constant corrections account for adjustments at the temporal boundaries.

The decoder cannot distinguish these two realizations and therefore cannot correct both.
The same ambiguity also applies to the marginal decoder.
Hence,
\begin{equation}
    \begin{aligned}
        t_{\rm E}\bnb  &\sim \frac{r}{2},\\
        t_{\rm E}\marg &\sim \frac{r}{2}.
    \end{aligned}
    \qquad \text{(stability)}
    \label{eq:walking-stability-te}
\end{equation}

\section{Transversal CNOT scaling}
\label{app:tcnot-fault-distance}

In this section, we extend the decoding graph analysis to tCNOT experiments under the skip-gate leakage model.
We first consider minimum-weight undetectable errors in memory and stability tCNOT experiments.
Next, we consider minimum-weight uncorrectable errors in memory and stability tCNOT experiments.
The final subsection presents numerical verification.

\subsection{Minimum-weight undetectable errors}

\subsubsection{Memory tCNOT}
\label{app:tcnot-memory}

For each patch in a memory tCNOT experiment, we first consider leakage errors sufficiently far from the tCNOT layer.
Their detector error patterns are identical to those in a single-patch memory experiment: each leakage error contributes at most one space-like edge along a shortest logical string.
We can therefore construct a length-$d$ logical path connecting the two open spatial boundaries of either patch, with all contributing leakage errors away from the tCNOT layer.
These paths give the same upper-bound construction for the control and target patches as in the single-patch memory experiment.

Leakage errors near the tCNOT layer can instead produce hyperedges involving detectors in both patches.
We assess whether these patterns reduce the fault distance by enumerating leakage configurations for small-distance codes.
The enumeration (\cref{tab:tcnot-memory-enumeration}) confirms that these hyperedges do not reduce the observed fault-distance scaling.
Together with the constructions away from the gate, these results support the same asymptotic scaling for both the control ($c$) and target ($t$) patches:
\begin{equation}
    d_{{\rm L},c}^{\mathrm{mem}} \sim d,
    \qquad
    d_{{\rm L},t}^{\mathrm{mem}} \sim d.
    \label{eq:tcnot-memory-dl}
\end{equation}

\subsubsection{Stability tCNOT}
\label{app:tcnot-stability}

\begin{figure}
    \centering
    \includegraphics[width=\columnwidth]{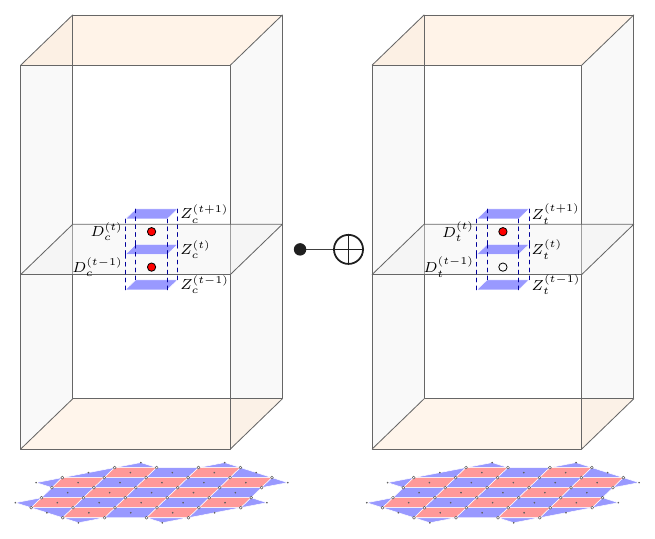}
    \caption{
        {\bf Spacetime schematic of a stability tCNOT experiment.}
        The left and right volumes represent the control ($c$) and target ($t$) blocks.
        The highlighted $Z$-check layers and the intervening detectors show the local structure near the tCNOT layer.
        Filled red circles mark the three detectors triggered by a measurement error in $Z_c^{(t)}$.
    }
    \label{fig:tcnot-stability-schematic}
\end{figure}

\cref{fig:tcnot-stability-schematic} illustrates the spacetime geometry and local $Z$-check structure near the tCNOT layer of the stability tCNOT experiment.
For the geometry with perfect $Z$ space boundaries shown here, we need only consider $Z$ detectors.
Let $r_1$ and $r_2$ denote the numbers of rounds before and after the tCNOT gates, respectively, with $r=r_1+r_2$ the total number of rounds.

The detector definitions near the tCNOT layer differ from those elsewhere in the circuit.
Write $m_p^{(s)}\in\{0,1\}$ for the measurement outcome of $Z_p^{(s)}$, where $p=c,t$ labels the patch and $s$ labels the round.
We assume that the tCNOT gates are applied after the stabilizer measurements in round $t$.
The detectors $D_c^{(t-1)}$, $D_c^{(t)}$, and $D_t^{(t-1)}$ keep the conventional definition as the parity of two consecutive outcomes of the same check.
In contrast, $D_t^{(t)}$ includes a control-patch outcome:
\begin{equation}
    \begin{aligned}
        D_c^{(t-1)} &= m_c^{(t-1)}\oplus m_c^{(t)},\\
        D_c^{(t)} &= m_c^{(t)}\oplus m_c^{(t+1)},\\
        D_t^{(t-1)} &= m_t^{(t-1)}\oplus m_t^{(t)},\\
        D_t^{(t)} &= m_c^{(t)}\oplus m_t^{(t)}\oplus m_t^{(t+1)}.
    \end{aligned}
    \label{eq:tcnot-stability-detectors}
\end{equation}

A measurement error in $Z_c^{(t)}$ flips $m_c^{(t)}$ and therefore triggers $D_c^{(t-1)}$, $D_c^{(t)}$, and $D_t^{(t)}$, as indicated by the filled red circles in \cref{fig:tcnot-stability-schematic}.
Thus a single measurement error near the tCNOT layer produces a hyperedge.

We first count measurement errors to identify the shortest undetectable logical error configurations.
In the target patch, a string of $r$ consecutive measurement errors that connects the two open time boundaries is an undetectable logical error.
A corresponding string of $r$ measurement errors in the control patch cancels all control-patch detectors but leaves $D_t^{(t)}$ triggered through the flipped outcome $m_c^{(t)}$.
The control string alone is therefore detectable.
To cancel this remaining detector, an additional target-patch string must connect $D_t^{(t)}$ to either the lower or upper open time boundary.
The two choices require $r_1$ or $r_2$ additional measurement errors respectively, giving a minimum additional temporal extent of $\min(r_1,r_2)$.

These temporal lengths also determine the leakage fault distances.
As in the preceding analysis, each leakage error away from the tCNOT layer contributes at most one elementary time-like edge.
In the moonwalking circuit, we observe that a leakage error near the tCNOT layer can contribute a weight-2 time-like advance, because of the additional tCNOT gates added to the circuit.
However, such contributions are confined to a region of limited temporal area around the gate and change the minimum leakage count by only $O(1)$.
Consequently, the target and control fault distances scale as
\begin{equation}
    \begin{aligned}
        d_{{\rm L},t}^{\mathrm{stab}} &\sim r,\\
        d_{{\rm L},c}^{\mathrm{stab}} &\sim r+\min(r_1,r_2),
    \end{aligned}
    \label{eq:tcnot-stability-dl}
\end{equation}
where $r=r_1+r_2$.
For stability tCNOT experiments on patches with perfect $X$ space boundaries, we consider only $X$ detectors, so the control and target fault-distance scalings are interchanged, with $d_{{\rm L},c}^{\mathrm{stab}}\sim r$ and $d_{{\rm L},t}^{\mathrm{stab}}\sim r+\min(r_1,r_2)$.

\subsection{Minimum-weight uncorrectable errors}

\subsubsection{Memory tCNOT}
Away from the tCNOT layer, each patch has the detector error patterns of a single-patch memory experiment, so the constructions of \cref{app:scaling} carry over to both patches.
Enumeration of the $d=3$ circuits (\cref{tab:tcnot-memory-enumeration}) shows that hyperedges near the tCNOT layer do not lower these values, supporting $t_{\rm E}\bnb \sim d$ and $t_{\rm E}\marg \sim 2d/3$ for both the control and target patches.

\subsubsection{Stability tCNOT}
\label{app:tcnot-stability-te}

We now consider decoding stability tCNOT experiments while simultaneously tracking the logical observables of the control and target patches.
A decoding failure occurs if either observable is decoded incorrectly.
As mentioned above, errors near the tCNOT layer can trigger more than two detectors, giving rise to a hypergraph decoding problem.
We therefore use the Tesseract decoder~\cite{beni2025tesseract} as our backend decoder.
Nevertheless, the preceding analysis of the residual mismatch $C\oplus E$ continues to apply: a failure requires the physical error $E$ and the correction $C$ to differ by a nontrivial logical operator.

In the foliated circuit, a mismatch between the actual error and a correction chosen from the candidate set for a single detected leakage error contributes at most one elementary time-like edge along a temporal logical path.
This bound also holds when the marginal decoder selects multiple candidates for the same leakage error.
The preceding analysis gives $d_{{\rm L},t}^{\mathrm{stab}}\sim r$ [\cref{eq:tcnot-stability-dl}], so a nontrivial logical path must span at least $r$ elementary time-like steps to leading order.
As argued above, the modified detector patterns near the tCNOT layer affect only a region of limited temporal area and therefore change the required number of leakage errors by at most $O(1)$.
Thus, at least $r-O(1)$ leakage errors are needed for $C\oplus E$ to form a logical path.
A time-like logical string in the target patch attains this bound up to $O(1)$ corrections, giving
\begin{equation}
    \begin{aligned}
        t_{\rm E}\marg &= r+O(1)\sim r,\\
        t_{\rm E}\bnb  &= r+O(1)\sim r.
    \end{aligned}
    \qquad \text{(foliated)}.
    \label{eq:tcnot-stability-te-foliated}
\end{equation}

For the moonwalking circuit, however, a single leakage error can produce a residual mismatch that advances a temporal logical path by at most two elementary steps, even when the correction respects the mutual exclusivity of the candidate leakage times.
A logical failure therefore requires at least $r/2-O(1)$ leakage errors.
Concatenating these weight-2 time-like mismatches yields a logical path with $r/2+O(1)$ leakage errors.
As in \cref{app:walking-stability}, the two corresponding error realizations have the same detected leakage errors and detector syndrome but different logical outcomes, so the decoder cannot correct both.
The tCNOT layer changes this construction by only $O(1)$ leakage errors.
Consequently, branch-and-bound decoding gives only
\begin{equation}
    t_{\rm E}\bnb=\frac{r}{2}+O(1)\sim\frac{r}{2}
    \qquad \text{(moonwalking)}.
    \label{eq:tcnot-stability-te-moonwalking}
\end{equation}

\subsection{Numerical verification}
\label{app:tcnot-numerical-verification}

We test the preceding analysis by enumerating leakage configurations in memory and stability tCNOT circuits at small code distances and round counts.
For each circuit, we determine the minimum-weight undetectable error $d_{\rm L}$ and the minimum-weight uncorrectable error $t_{\rm E}$ for both the marginal and branch-and-bound decoders.
Both circuit implementations use two $X$-memory patches in the memory tCNOT experiments.
The stability tCNOT experiments use two $X$-stability patches for moonwalking and two $Z$-stability patches for foliation.
For each patch, the QEC intervals before and after the tCNOT have $r_1$ and $r_2$ rounds, respectively.
In walking-type circuits, we measure both the $Z$ and $X$ checks $r_1$ times before the gate and $r_2$ times after it.
For the foliated circuit, we decompose the tCNOT as $H_t\,\mathrm{tCZ}\,H_t$, where tCZ denotes a transversal controlled-$Z$ gate and teleportation of the target qubits implements the Hadamard layers $H_t$.
On the control patch, we measure each check type $r_1$ times before the tCZ layer.
After it, we measure the $Z$ checks $r_2+1$ times and the $X$ checks $r_2$ times.
On the target patch, we measure each check type $r_1+1$ times before the tCZ layer, and after it, we measure the $Z$ checks $r_2+1$ times and the $X$ checks $r_2$ times.
\cref{tab:tcnot-memory-enumeration,tab:tcnot-stability-enumeration} summarize the memory and stability tCNOT results respectively. We define the total number of rounds as $r_m=r_1+r_2$ for moonwalking and $r_f=r_1+r_2+2$ for foliation. The offset of two in  $r_f$ results from the check schedule described above.

\begin{table}
    \centering
    \setlength{\tabcolsep}{4pt}
    \begin{tabular*}{\columnwidth}{@{\extracolsep{\fill}}lcccc}
        \toprule
        Circuit & $d$ & $d_{\rm L}$ & $t_{\rm E}\marg$ & $t_{\rm E}\bnb$ \\
        \midrule
        Moonwalking & 3 & 3 & 2 & 3 \\
        Foliated    & 3 & 3 & 2 & 3 \\
        \bottomrule
    \end{tabular*}
    \caption{
        {\bf Memory tCNOT enumeration.} Leakage enumeration results for memory tCNOT experiments with two $X$-memory patches.
        }
    \label{tab:tcnot-memory-enumeration}
\end{table}

\begin{table}
    \centering
    \setlength{\tabcolsep}{4pt}
    \renewcommand{\arraystretch}{1.15}
    \begin{tabular*}{\columnwidth}{@{\extracolsep{\fill}}lccccc@{}}
        \toprule
        \multirow{2}{*}{Circuit} & \multirow{2}{*}{$(r_1,r_2)$}
        & \multicolumn{2}{c}{Control} & \multicolumn{2}{c}{Target} \\
        \cmidrule(lr){3-4} \cmidrule(l){5-6}
        & & $d_{\rm L}$ & $t_{\rm E}$ & $d_{\rm L}$ & $t_{\rm E}$ \\
        \midrule
        \multirow{2}{*}{Moonwalking} & $(1,2)$ & 2 & 2 & 4 & 3 \\
                                    & $(3,1)$ & 3 & 2 & 5 & 3 \\
        \midrule
        \multirow{3}{*}{Foliated}   & $(1,0)$ & 3 & 3 & 3 & 3 \\  
                                    & $(2,0)$ & 4 & 4 & 4 & 4 \\
                                    & $(1,1)$ & 5 & 5 & 4 & 4 \\
        \bottomrule
    \end{tabular*}    
    \caption{
        {\bf Stability tCNOT enumeration.} Leakage enumeration results for stability tCNOT experiments with code distance $d=2$.
        Both patches are $X$-stability patches for moonwalking and $Z$-stability patches for foliation.
        Results are reported separately for the control and target patches.
        In all cases shown, $t_{\rm E}=t_{\rm E}\bnb=t_{\rm E}\marg$.
    }
    \label{tab:tcnot-stability-enumeration}
\end{table}

For the memory tCNOT experiments, enumeration of the $d=3$ circuits gives the same results as the corresponding memory experiments without a tCNOT layer (\cref{tab:tcnot-memory-enumeration}).
For both moonwalking and foliation, the branch-and-bound decoder achieves the optimal distance, $t_{\rm E}\bnb=d_{\rm L}=3$, whereas the marginal decoder gives $t_{\rm E}\marg=2$.
These values remain unchanged across the different $(r_1,r_2)$ settings we tested.
We also obtain the same values when using an $X$-memory control patch and a $Z$-memory target patch.

\begin{figure}
    \centering
    \subfloat{%
        \includegraphics[width=0.88\columnwidth]{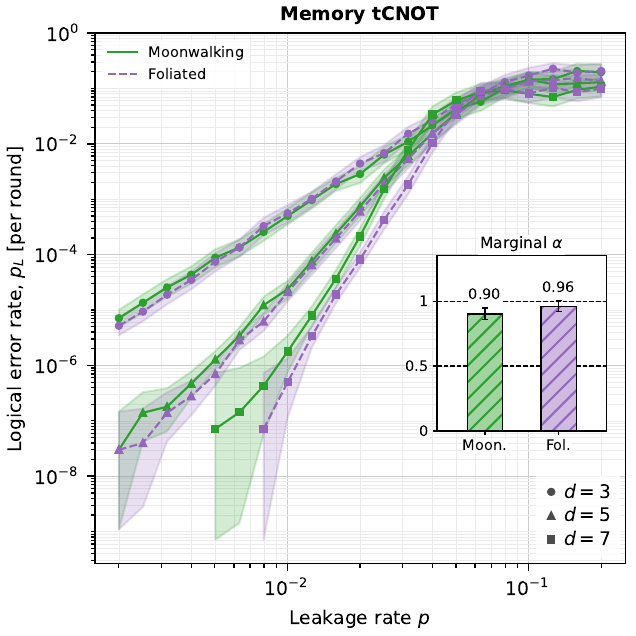}%
        \llap{\makebox[0.88\columnwidth][l]{\hspace{0.12\columnwidth}%
            \raisebox{\dimexpr0.88\columnwidth-0.65\baselineskip\relax}[0pt][0pt]{\sffamily\bfseries (a)}}}%
        \label{subfig:tcnot-logical-error-rates:memory}%
    }\par\smallskip
    \subfloat{%
        \includegraphics[width=0.88\columnwidth]{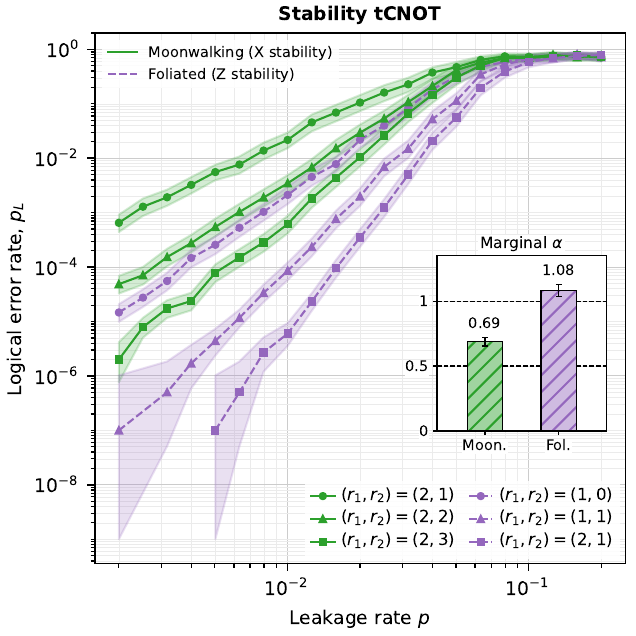}%
        \llap{\makebox[0.88\columnwidth][l]{\hspace{0.12\columnwidth}%
            \raisebox{\dimexpr0.88\columnwidth-0.65\baselineskip\relax}[0pt][0pt]{\sffamily\bfseries (b)}}}%
        \label{subfig:tcnot-logical-error-rates:stability}%
    }
    \caption{\textbf{Logical error rates of memory and stability tCNOT experiments under skip-gate leakage.}
    \textbf{(a)} Logical error rate per round for $X$-memory tCNOTs at code distances $d=3,5,7$.
    \textbf{(b)} Failure probability $P_{\rm fail}$ per shot for moonwalking $X$-stability and foliated $Z$-stability tCNOTs.
    The legend gives the numbers of rounds before and after the tCNOT, $(r_1,r_2)$, for each circuit.
    The insets show the fitted subthreshold scaling exponents $\alpha$ for marginal decoding.}
    \label{fig:tcnot-logical-error-rates}
\end{figure}

For the moonwalking stability tCNOT experiments, we expect $d_{{\rm L},c}^{\mathrm{stab}}\sim r$ and $d_{{\rm L},t}^{\mathrm{stab}}\sim r+\min(r_1,r_2)$.
For the configurations in \cref{tab:tcnot-stability-enumeration}, we find
\begin{equation}
    \begin{aligned}
        d_{{\rm L},c}^{\mathrm{stab}} &= r_m-1,\\
        d_{{\rm L},t}^{\mathrm{stab}} &= r_m+\min(r_1,r_2),
    \end{aligned}
\end{equation}
consistent with these scalings.
We attribute the one-unit reduction in the control-patch fault distance to leakage errors near the tCNOT layer, where the additional transversal gates allow a single leakage error to create a weight-2 time-like edge.
The decoding results are also consistent with our analysis:
\begin{equation}
    t_{{\rm E},p}\marg=t_{{\rm E},p}\bnb\sim d_{{\rm L},p}^{\mathrm{stab}}/2,
    \qquad p\in\{c,t\}.
\end{equation}

For foliation, we expect the target and control fault distances to scale as $r$ and $r+\min(r_1,r_2)$, respectively.
We find
\begin{equation}
    \begin{aligned}
        d_{{\rm L},t}^{\mathrm{stab}} &= r_f,\\
        d_{{\rm L},c}^{\mathrm{stab}} &= r_f+\min(r_1,r_2).
    \end{aligned}
\end{equation}
Both the marginal and branch-and-bound decoders attain the optimal fault distance on each patch:
\begin{equation}
    t_{{\rm E},p}\marg=t_{{\rm E},p}\bnb=d_{{\rm L},p}^{\mathrm{stab}},
    \qquad p\in\{c,t\}.
\end{equation}

\Cref{fig:tcnot-logical-error-rates} compares the logical error rates of moonwalking and foliated tCNOT circuits as a function of the leakage rate $p$.
We consider CZ gate leakage only.
We use the marginal decoder with Tesseract's long-beam configuration as the backend~\cite{beni2025tesseract}.
For memory tCNOTs, both implementations use two distance-$d$ $X$-memory patches with $r_1=r_2=d$.
For stability tCNOTs, the moonwalking circuit uses two $d=6$ $X$-stability patches and the foliated circuit uses two $d=6$ $Z$-stability patches.
We use the compact gate ordering for the foliated circuit in both memory and stability tCNOTs.
The legend lists the $(r_1,r_2)$ values for each circuit.
In all experiments shown, we track the logical observables of both the control and target patches and count a shot as a logical failure if at least one of the two observables is flipped after error correction.
The two implementations yield comparable logical error rates and scaling for memory tCNOTs.
For stability tCNOTs, the foliated circuit has better subthreshold scaling than moonwalking and achieves substantially lower logical error rates, in agreement with the preceding analysis.

\section{Neutral-atom simulations}
\label{app:harvard_noise_model}
\subsection{Noise model}
\begin{table*}
    \centering
    \setlength{\tabcolsep}{8pt}
    \renewcommand{\arraystretch}{1.2}
    \begin{tabular*}{\textwidth}{@{\extracolsep{\fill}}llccc}
        \toprule
        \textbf{Mechanism} $m$ & \textbf{Nominal Pauli rates} $e_{\mathrm{P},0}$ & \textbf{Loss} $e_{\mathrm{L},0}$ & $(x_{\mathrm{P}}, x_{\mathrm{L}})$ & \textbf{Infidelity} $f_{\mathrm{tot},0}$ \\
        \midrule
        SLM idle ($\mu\mathrm{s}^{-1}$) & $(p_X, p_Y, p_Z) = (0.1, 0.1, 0.5)\times 10^{-6}$ & $0.09\times 10^{-6}$ & $(2/3,\, 1)$ & $0.557\times 10^{-6}$ \\
        AOD idle ($\mu\mathrm{s}^{-1}$) & $(p_X, p_Y, p_Z) = (0.5, 0.5, 1.0)\times 10^{-6}$ & $0.09\times 10^{-6}$ & $(2/3,\, 1)$ & $1.423\times 10^{-6}$ \\
        Two-qubit $\mathrm{CZ}$ gate & \begin{tabular}[t]{@{}l@{}}$IX,IY,XI,YI = 5\times 10^{-5}$;\\ $IZ,ZI = 5\times 10^{-4}$;\; $ZZ = 10^{-3}$\end{tabular} & $1.83\times 10^{-3}$ & $(4/5,\, 2)$ & $5.420\times 10^{-3}$ \\
        $\mathrm{CZ}$ spectator (EZ) & $(p_X, p_Y, p_Z) = (0.05, 0.05, 1.0)\times 10^{-3}$ & $1.83\times 10^{-3}$ & $(2/3,\, 1)$ & $2.563\times 10^{-3}$ \\
        Hadamard ($H$) & $(p_X, p_Y, p_Z) = (0.2, 0.6, 0.2)\times 10^{-3}$ & $0$ & $(2/3,\, 1)$ & $0.667\times 10^{-3}$ \\
        Reset ($R$) & $p_X = 0.5\times 10^{-3}$ & $0.60\times 10^{-3}$ & $(2/3,\, 1)$ & $0.933\times 10^{-3}$ \\
        Measurement ($M$) & $p_X = 2.4\times 10^{-3}$ & $2.40\times 10^{-3}$ & $(2/3,\, 1)$ & $4.000\times 10^{-3}$ \\
        Handoff & $0$ & $7.32\times 10^{-3}$ & $(2/3,\, 1)$ & $7.320\times 10^{-3}$ \\
        \bottomrule
    \end{tabular*}

    \caption{{\bf Neutral-atom noise parameters.} Neutral-atom noise model parameters at nominal scale ($p = 1.0, \lambda = 0.5$) and state-averaged infidelity weights $(x_{\mathrm{P}}, x_{\mathrm{L}})$.
    For the two-qubit $\mathrm{CZ}$ gate, $e_{\mathrm{L},0}$ denotes the per-atom loss probability (half of the total two-qubit gate loss, so $x_{\mathrm{L}} = 2$).
    Continuous idling rates are in $\mu\mathrm{s}^{-1}$, with inter-zone shuttling duration $T_{\mathrm{shuttle}} = 1285\,\mu\mathrm{s}$ and intra-zone rearrangement duration $T_{\mathrm{CZ,idle}} = 470\,\mu\mathrm{s}$.}
    \label{tab:harvard_ballpark_params}
\end{table*}
We model a zoned neutral-atom processor in which atoms reside in static SLM traps by default and are handed off into a single mobile AOD array whenever shuttled.
All nominal parameters ($p = 1.0, \lambda = 0.5$) are listed in \cref{tab:harvard_ballpark_params}, where continuous idling rates $\gamma$ acting over duration $\Delta t$ yield discrete probabilities $1 - (1 - \gamma)^{\Delta t}$.
Each round includes:
\begin{enumerate}
    \item Reset and measurement errors: each reset is followed by a bit-flip ($X$) error and atom loss, and each measurement is preceded by a bit-flip ($X$) error and atom loss, applied identically to data and ancilla qubits.
    
    \item Inter-zone shuttling: after each reset and before each measurement, the reset or measured atoms shuttle in the AOD for $T_{\mathrm{shuttle}} = 1285\,\mu\mathrm{s}$, while all other atoms idle in SLM traps for $T_{\mathrm{shuttle}}$.
    
    \item Intra-zone rearrangement: between consecutive physical $\mathrm{CZ}$ layers within an entangling block, mobile atoms idle in AOD traps and stationary atoms idle in SLM traps for $T_{\mathrm{CZ,idle}} = 470\,\mu\mathrm{s}$.
    On the rotated lattice, a single AOD moves $X$- and $Z$-ancilla subgrids sequentially, splitting the four stabilizer steps of walking and moonwalking circuits into eight $\mathrm{CZ}$ layers (seven $T_{\mathrm{CZ,idle}}$ intervals) per round, whereas on the unrotated lattice all ancillas share a single bipartite subgrid and move together across four $\mathrm{CZ}$ layers (three $T_{\mathrm{CZ,idle}}$ intervals) per round; in foliation, each half-round has four stabilizer $\mathrm{CZ}$ layers (three $T_{\mathrm{CZ,idle}}$ intervals) followed by an inter-layer teleportation $\mathrm{CZ}$ layer.
    
    \item Handoffs: each transfer between SLM and AOD traps incurs atom loss equal to twice the total two-qubit $\mathrm{CZ}$ gate loss ($4\times 1.83\times 10^{-3} = 7.32\times 10^{-3}$ per transferred atom at nominal scale).
    
    \item Gate and spectator errors: each Hadamard gate is followed by a single-qubit Pauli channel and atom loss.
    During each physical $\mathrm{CZ}$ layer, gated pairs undergo a two-qubit Pauli channel and independent per-atom loss (half of the total two-qubit $\mathrm{CZ}$ gate loss per atom), while all non-interacting atoms undergo spectator entangling-zone Pauli and loss errors.
    In walking and moonwalking circuits, one $\mathrm{CZ}$ layer of the role-exchange step is absorbed into state preparation or classical post-processing (leaving four physical $\mathrm{CZ}$ steps per round), and surviving remapped atoms are returned to their canonical lattice sites after measurement.
\end{enumerate}
For each mechanism with nominal Pauli and loss infidelities $f_{\mathrm{P}} = x_{\mathrm{P}}\sum e_{\mathrm{P},0}$, $f_{\mathrm{L}} = x_{\mathrm{L}}e_{\mathrm{L},0}$, and $f_{\mathrm{tot},0} = f_{\mathrm{P}} + f_{\mathrm{L}}$, we rescale the rates by an overall multiplier $p \ge 0$ and loss fraction $\lambda \in [0,1]$ at fixed total infidelity $p\,f_{\mathrm{tot},0}$.
For mechanisms with both $f_{\mathrm{P}}, f_{\mathrm{L}} > 0$,
\begin{align}
    e_{\mathrm{L}}(p,\lambda) &= p\,e_{\mathrm{L},0}\,\frac{\lambda f_{\mathrm{tot},0}}{\lambda f_{\mathrm{L}} + (1-\lambda)f_{\mathrm{P}}}, \nonumber\\
    e_{\mathrm{P}}(p,\lambda) &= p\,e_{\mathrm{P},0}\,\frac{(1-\lambda) f_{\mathrm{tot},0}}{\lambda f_{\mathrm{L}} + (1-\lambda)f_{\mathrm{P}}}.
\end{align}
For Hadamard gates ($f_{\mathrm{L}} = 0$), we set $e_{\mathrm{P}}(p,\lambda) = p(1-\lambda)e_{\mathrm{P},0}$ and $e_{\mathrm{L}}(p,\lambda) = p\lambda f_{\mathrm{tot},0}/x_{\mathrm{L}}$; for handoffs ($f_{\mathrm{P}} = 0$), we set $e_{\mathrm{L}}(p,\lambda) = p\lambda e_{\mathrm{L},0}$ and $p_X = p_Y = p_Z = p(1-\lambda)f_{\mathrm{tot},0}/(3x_{\mathrm{P}})$.
Thus $\lambda = 0$, $0.5$, and $1$ correspond to pure Pauli noise, nominal experimental rates, and pure atom loss, respectively.

\subsection{Path-counting model}
\label{app:partition_function}

\begin{figure}
    \centering
    \includegraphics[width=\columnwidth]{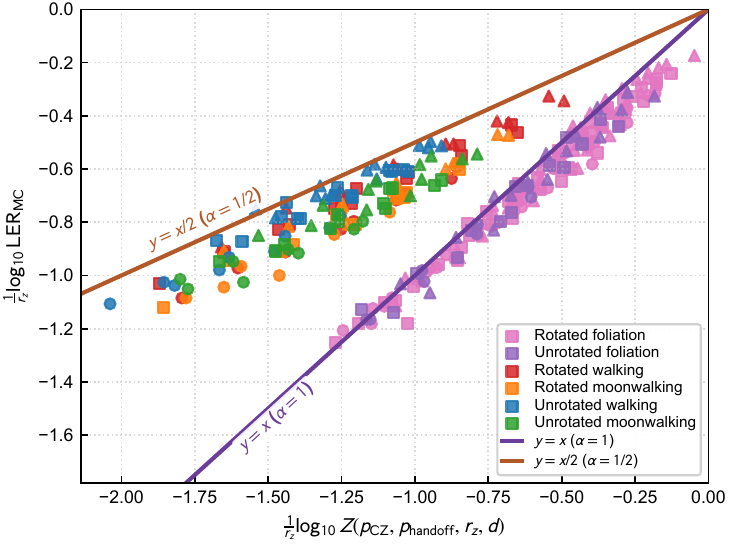}
    \caption{\textbf{Percolation partition function $Z$ versus Monte Carlo logical error rate.}
    Per-round log logical error rate $\frac{1}{r_z}\log_{10}\mathrm{LER}_{\mathrm{MC}}$ versus $\frac{1}{r_z}\log_{10} Z(p_{\mathrm{CZ}}, p_{\mathrm{handoff}}, r_z, d)$ across all six protocols for $d \in \{2, 4, 6\}$ (circles, squares, triangles), $r_z \in \{4,\dots,7\}$, and $p_{\mathrm{handoff}}/p_{\mathrm{CZ}} \in \{0, 2\}$ (Pearson correlation $0.98$ for foliation and $0.94$ for walking and moonwalking).
    Solid lines show the zero-offset rays $y = x$ ($\alpha = 1$, foliation) and $y = x/2$ ($\alpha = 1/2$, walking and moonwalking) from the origin.}
    \label{fig:appendix_Z_correlation}
\end{figure}

Here $r_z$ counts stabilizer rounds, so a stability circuit has $r = r_z + 1$ layers of $Z$ checks in the notation of \cref{sec:experiments}.
We present a path-counting model to explain two features of the stability simulations: the $\alpha = 0.5$ scaling exponent of walking and moonwalking circuits under pure $\mathrm{CZ}$ loss, and the sensitivity of rotated walking to handoff loss.
We estimate the multiplicity of undetectable heralded errors via a transfer-matrix partition function $Z$ that sums over directed walks between the open time boundaries of the $Z$-stability decoding graph.
Because $Z$ ignores self-avoidance and decoder tie-breaking weights, it is an entropic proxy rather than an exact decoder calculation.
Below, we verify that $Z$ tracks the Monte Carlo logical error rate (\cref{fig:appendix_Z_correlation}) and use its largest eigenvalue to compare the threshold and handoff conversion across protocols (\cref{tab:kappa_summary}).

\subsubsection{Partition function \texorpdfstring{$Z$}{Z} and comparison with Monte Carlo}

A stability circuit with $r_z$ rounds has $r_z + 1$ layers of spatial $Z$ checks between its open initial and final time boundaries.
A single atom loss generates candidate detector-graph edges with round displacement $\Delta t \in \{0, 1, 2\}$: equal-time hops within a round ($\Delta t = 0$, matrix $W_0$), one-round transitions from $t - 1$ to $t$ ($\Delta t = 1$, matrix $T_1$), and two-round transitions from $t - 2$ to $t$ ($\Delta t = 2$, matrix $T_2$).
Both $T_1$ and $T_2$ include vertical steps on the same check (diagonal entries) and space-time diagonal hops to neighboring checks (off-diagonal entries).
Summing over all directed paths across the $r_z + 1$ layers defines the partition function
\begin{equation}
    Z = \begin{pmatrix} \mathbf{1}^T & \mathbf{0}^T \end{pmatrix}
    \begin{pmatrix} W_0 + T_1 & T_2 \\ I & 0 \end{pmatrix}^{\!r_z + 1}
    \begin{pmatrix} \mathbf{1} \\ \mathbf{0} \end{pmatrix},
    \label{eq:Z_3D_def}
\end{equation}
where $\mathbf{1}$ is the all-ones boundary vector and the identity block $I = \mathcal{O}(p_{\mathrm{CZ}}^0)$ carries the previous layer across a two-round step at zero loss cost.

\Cref{eq:Z_3D_def} explains the two scaling exponents in \cref{fig:appendix_Z_correlation}.
In \emph{foliation}, qubits are measured and reset every half-round ($T_2 = 0$), so \cref{eq:Z_3D_def} reduces to $\mathbf{1}^T(W_0 + T_1)^{r_z+1}\mathbf{1} = \mathcal{O}(p_{\mathrm{CZ}}^{r_z + 1})$ ($\alpha = 1$, ray $y = x$).
In \emph{walking} and \emph{moonwalking}, each atom lives for two rounds and switches roles before readout ($T_2 \neq 0$, \cref{subsec:simulations:neutral-atom}); alternating $T_2 = \mathcal{O}(p_{\mathrm{CZ}})$ with the identity block $I = \mathcal{O}(p_{\mathrm{CZ}}^0)$ advances two layers per loss event, so only $\lceil r_z/2\rceil$ loss events are needed to span the circuit, yielding $\alpha = 1/2$ (ray $y = x/2$).
The vertical offset from the zero-offset rays in \cref{fig:appendix_Z_correlation} arises because $Z$ assigns unit weight to all candidate walks, whereas a physical loss causes a decoding failure only when its Pauli error flips the target $Z$ check (probability $1/2$) and outweighs the complementary path.

\subsubsection{Effective loss rate and handoff sensitivity}

For large $r_z$, \cref{eq:Z_3D_def} is dominated by the largest eigenvalue $\Lambda(d) = \Lambda_0(d) + (p_{\mathrm{handoff}}/p_{\mathrm{CZ}})\Lambda_H(d)$ of $W_0 + T_1 + T_2$ (with two-qubit gate loss rate $p_{\mathrm{CZ}}$, i.e., per-atom loss probability $p_{\mathrm{CZ}}/2$ before each physical $\mathrm{CZ}$ gate), where $\Lambda_0$ and $\Lambda_H$ are the $\mathrm{CZ}$ and handoff path multiplicities per step.
This gives
\begin{equation}
    Z \propto \left(\frac{p_{\mathrm{eff}}}{p_{\mathrm{th,CZ}}}\right)^{\!r_z + 1}, \qquad p_{\mathrm{eff}} = p_{\mathrm{CZ}} + \kappa(d)\,p_{\mathrm{handoff}},
    \label{eq:Z_spectral_peff_pth}
\end{equation}
with pure-$\mathrm{CZ}$ threshold $p_{\mathrm{th,CZ}} \propto 1/\Lambda_0(d)$ (calibrated to the bulk foliation threshold $p_{\mathrm{th,CZ}}(\infty) = 0.135$) and handoff conversion coefficient $\kappa(d) = \Lambda_H(d)/\Lambda_0(d)$.

\begin{table}
    \centering
    \setlength{\tabcolsep}{4pt}
    \begin{tabular*}{\columnwidth}{@{\extracolsep{\fill}}lcccc}
        \toprule
        \textbf{Protocol} & $\Lambda_0(\infty)$ & $\kappa(\infty)$ & $\kappa(4)$ & $\kappa_{\mathrm{MC}}(4)$ \\
        \midrule
        Rot.\ foliation     & $20$ & $0.70$ & $0.80$ & $0.95$ \\
        Unrot.\ foliation   & $20$ & $0.70$ & $0.74$ & $0.78$ \\
        Unrot.\ walking     & $24$ & $1.12$ & $1.22$ & $0.98$ \\
        Unrot.\ moonwalking & $20$ & $2.00$ & $2.07$ & $1.72$ \\
        Rot.\ moonwalking   & $20$ & $3.20$ & $3.59$ & $3.11$ \\
        Rot.\ walking       & $24$ & $3.71$ & $4.44$ & $4.49$ \\
        \bottomrule
    \end{tabular*}

    \caption{{\bf Path multiplicities and handoff conversion.} Bulk $\mathrm{CZ}$ loss multiplicity $\Lambda_0(\infty)$ (in per-gate $p_{\mathrm{CZ}}$ units; per-atom multiplicities are $2\Lambda_0(\infty) \in \{40, 48\}$) and handoff conversion coefficient $\kappa = \Lambda_H / \Lambda_0$ at $d \to \infty$, analytical $d = 4$, and extracted from Monte Carlo at $d = 4$ under skip-gate ($\mathrm{loss\text{-}before\text{-}CZ}$) timing.}
    \label{tab:kappa_summary}
\end{table}

\Cref{tab:kappa_summary} compares $\Lambda_0$ and $\kappa$ across all six protocols.
Under skip-gate $\mathrm{CZ}$ loss, an ancilla lost during a walking round skips its end-of-round role swap and stays on the same check in the next round, raising $\Lambda_0(\infty)$ from $20$ (moonwalking and foliation) to $24$ and lowering $p_{\mathrm{th,CZ}}$.
Conversely, moonwalking performs its role-swap handoffs at the start of each round before all four stabilizer $\mathrm{CZ}$s, increasing $\kappa$ relative to unrotated walking.
On the rotated lattice, moving the $X$- and $Z$-ancilla subgrids sequentially roughly triples $\kappa$ ($\kappa > 3$), explaining why handoff loss reverses the stability ordering of rotated and unrotated walking between the stability panels of \cref{subfig:loss_only_comparison:cz} and \cref{subfig:loss_only_comparison:handoff}.
Across all $18$ protocol-distance pairs ($d \in \{2, 4, 6\}$), the analytical values $\kappa(d)$ track $\kappa_{\mathrm{MC}}(d)$ extracted from the horizontal shift of the Monte Carlo curves between $p_{\mathrm{handoff}} = 0$ and $p_{\mathrm{handoff}} = 2p_{\mathrm{CZ}}$ ($R^2 = 0.93$).

\end{document}